\pdfoutput=1

\documentclass[11pt,twoside,a4paper,cmspaper,final,collab]{cms-tdr}
\def\svnVersion{227786}\def\cmsCernNo{2013-213}\def\cmsCernDate{\today}\def\cmsMessage{Published in the Journal of High Energy Physics as \href{http://dx.doi.org/10.1007/JHEP01(2014)163}{\doi{10.1007/JHEP01(2014)163.}}}

\begin{document}\cmsNoteHeader{SUS-13-013}

\hyphenation{had-ron-i-za-tion}
\hyphenation{cal-or-i-me-ter}
\hyphenation{de-vices}

\RCS$Revision: 218406 $
\RCS$HeadURL: svn+ssh://svn.cern.ch/reps/tdr2/papers/SUS-13-013/trunk/SUS-13-013.tex $
\RCS$Id: SUS-13-013.tex 218406 2013-11-27 16:17:09Z alverson $
\newcommand{\fullLumi}{19.49\fbinv}  

\newcommand{\chargino}{\ensuremath{\widetilde{\chi}_{1}^{\pm}}\xspace}
\newcommand{\neutralino}{\ensuremath{\widetilde{\chi}_{2}^{0}}\xspace}
\newcommand{\lsp}{\ensuremath{\widetilde{\chi}_{1}^{0}}\xspace}
\newcommand{\slep}{\ensuremath{\widetilde{\ell}}\xspace}
\newcommand{\xslep}{\ensuremath{x_{\slep}}\xspace}
\newcommand{\snu}{\ensuremath{\widetilde{\nu}}\xspace}
\newcommand{\mchi}{\ensuremath{{m_{\chiz_2}=m_{\chipm_1}}}\xspace}
\newcommand{\Irel}{\ensuremath{I_\text{rel}}\xspace} 
\newcommand{\W}{\ensuremath{\cmsSymbolFace{W}}\xspace} 
\newcommand{\ZZ}{\ensuremath{\cmsSymbolFace{ZZ}}\xspace}
\newcommand{\udsgjet}{\ensuremath{\cPqu\cPqd\cPqs\cPg\text{-jet}}\xspace}
\newcommand{\cjet}{\ensuremath{\cPqc\text{-jet}}\xspace}
\newcommand{\bjet}{\ensuremath{\cPqb\text{-jet}}\xspace}
\newcommand{\bjets}{\ensuremath{\cPqb\text{-jets}}\xspace}
\newcommand{\MT}{\ensuremath{M_\mathrm{T}}\xspace}
\newcommand{\mdil}{\ensuremath{M_{\ell\ell}}\xspace}
\newcommand{\mjj}{\ensuremath{M_{jj}}\xspace}

\newcommand{\nbjets}{\ensuremath{N_{\text{b-jets}}}\xspace}

\newcommand{\njets}{\ensuremath{N_{\mathrm{jets}}}\xspace}
\newcommand{\ST}{\ensuremath{S_{\mathrm{T}}}\xspace}
\newcommand{\zjets}{\ensuremath{\Z+\text{jets}}\xspace}
\newcommand{\wjets}{\ensuremath{\W+\text{jets}}\xspace}
\newcommand{\wwjets}{\ensuremath{\W\W+\text{jets}}\xspace}
\newcommand{\gjets}{\ensuremath{\gamma+\text{jets}}\xspace}

\newcommand{\zzmet}{$\Z\Z+\MET$\xspace}
\newcommand{\wzmet}{$\W\Z+\MET$\xspace}
\newcommand{\wzzmet}{$\W\Z/\Z\Z+\MET$\xspace}
\newcommand{\cls}{CL$_\text{s}$\xspace}

\def\mrm{\mathrm}
\def\ra{\rightarrow}
\newcommand{\nn}{\nonumber \\ }

\newcommand{\at}{\symbol{64}}
\newcommand{\sye}[1]{\ensuremath{~\pm #1}}
\newcommand{\ase}[2]{\ensuremath{^{~+ #1}_{~- #2}}}
\newcommand{\asi}[2]{\ensuremath{^{~- #1}_{~+ #2}}}

\newcommand{\jp}{\ensuremath{J/\psi}}
\newcommand{\eess}{\ensuremath{ee}}
\newcommand{\mmss}{\ensuremath{\mu\mu}}
\newcommand{\emss}{\ensuremath{e\mu}}
\newcommand{\eepm}{\ensuremath{e^+ e^-}}
\newcommand{\mmpm}{\ensuremath{\mu^+ \mu^-}}
\newcommand{\ttpm}{\ensuremath{\tau^+ \tau^-}}
\newcommand{\empm}{\ensuremath{e^\pm \mu^\mp}}
\newcommand{\dy}{\ensuremath{Z/\gamma^\star}}
\newcommand{\dyee}{\ensuremath{\dy\to\eepm}}
\newcommand{\dymm}{\ensuremath{\dy\to\mmpm}}
\newcommand{\dytt}{\ensuremath{\dy\to\ttpm}}
\newcommand{\dyll}{\ensuremath{\dy\to\ell\ell}}
\newcommand{\wln}{\ensuremath{W\to \ell\nu_{\ell}}}
\newcommand{\wen}{\ensuremath{W\to e\nu_{e}}}
\newcommand{\wmn}{\ensuremath{W\to\mu\nu_{\mu}}}
\newcommand{\wtn}{\ensuremath{W\to\tau\nu_{\tau}}}
\newcommand{\wqq}{\ensuremath{W\to q\bar{q}'}}

\newcommand{\roots}{\ensuremath{\sqrt{s}}}
\newcommand{\lhcE}[1]{\ensuremath{\roots ={#1}\TeV}}
\newcommand{\eps}[1]{\ensuremath{\epsilon_{#1}}}
\newcommand{\pp}{\ensuremath{pp}}
\newcommand{\ttg}{\ensuremath{\ttbar\gamma}}
\newcommand{\ttw}{\ensuremath{\ttbar W}}
\newcommand{\ttz}{\ensuremath{\ttbar Z}}
\newcommand{\wgamma}{\ensuremath{W\gamma^{*}}}

\newcommand{\rphi}{\text{$r$-$\phi$}}
\newcommand{\etaphi}{\text{$\eta$-$\phi$}}
\newcommand{\rz}{\text{$r$-$z$}}
\newcommand{\aeta}{\ensuremath{\abs{\eta}}}
\newcommand{\Ht}{\ensuremath{H_T}}
\newcommand{\met} {\ensuremath{E\!\!\!\!/_T}}
\newcommand{\mt}{\ensuremath{m_T}}
\newcommand{\nbtags}{$\#$ b-tagged jets}
\newcommand{\cmsmet} {\ensuremath{\mathrm{E^{miss}_T}}}
\newcommand{\tcmet}{\ensuremath{\mrm{tcMET}}}
\newcommand{\jets}{\ensuremath{\mrm{jets}}}
\newcommand{\isotk}{\ensuremath{I_\mrm{trk}}}
\newcommand{\isocal}{\ensuremath{I_\mrm{cal}}}
\newcommand{\mll}{\ensuremath{M_{\ell\ell}}}
\newcommand{\mee}{\ensuremath{M_{ee}}}
\newcommand{\mmumu}{\ensuremath{M_{\mu\mu}}}
\newcommand{\gt}{\ensuremath{>}}
\newcommand{\lt}{\ensuremath{<}}
\newcommand{\absetacut}{\ensuremath{\abs{\eta} < 2.4}}
\newcommand{\dr}{\ensuremath{\Delta R}}
\newcommand{\relIso}{\ensuremath{Iso}}

\newcommand{\tnp}{tag $\rm{\&}$ probe}
\newcommand{\highpt}{high $p_{T}$}
\newcommand{\lowpt}{low $p_{T}$}
\newcommand{\vlowpt}{very low $p_{T}$}
\newcommand{\ttdilS}{\ensuremath{\ttbar\to\ell\ell X}}
\newcommand{\ttslbS}{\ensuremath{\ttbar\to\ell(b\to\ell) X}}
\newcommand{\ttsloS}{\ensuremath{\ttbar\to\ell(b\!\!\!/\to\ell) X}}
\newcommand{\SFnn}{\ensuremath{N^{\rm Wj,raw}_{nn}}}

\newlength\cmsFigWidth
\ifthenelse{\boolean{cms@external}}{\setlength\cmsFigWidth{0.85\columnwidth}}{\setlength\cmsFigWidth{0.4\textwidth}}
\ifthenelse{\boolean{cms@external}}{\providecommand{\cmsLeft}{top}}{\providecommand{\cmsLeft}{left}}
\ifthenelse{\boolean{cms@external}}{\providecommand{\cmsRight}{bottom}}{\providecommand{\cmsRight}{right}}

\newcommand{\ifPAS}[1]{\ifthenelse{\boolean{cms@pas}}{#1}{}}
\newcommand{\ifPAPER}[1]{\ifthenelse{\NOT\boolean{cms@pas}}{#1}{}}

\cmsNoteHeader{SUS-12-017} 
\title{Search for new physics in events with same-sign dileptons
and jets in pp collisions at $\sqrt{s}=8$\TeV}

\date{\today}

\abstract{
A search for new physics is performed based on events with jets and a pair of isolated, same-sign leptons.
The results are obtained using
a sample of proton-proton collision data collected by the CMS experiment at a centre-of-mass energy of 8\TeV at the LHC,
corresponding to an integrated luminosity of 19.5\fbinv.
In order to be sensitive to a wide variety of possible signals
beyond the standard model, multiple search regions
defined by the missing transverse energy, the hadronic energy, the number of jets and b-quark jets,
and the transverse momenta of the leptons in the events are considered.
No excess above the standard model background expectation is observed and
 constraints are set on a number of models for new physics,
 as well as on the same-sign top-quark pair and quadruple-top-quark production cross sections.
Information on event selection efficiencies is also provided, so that the
results can be used to confront an even broader class of new physics models.}

\hypersetup{%
pdfauthor={CMS Collaboration},%
pdftitle={Search for new physics in events with same-sign dileptons
and jets in pp collisions at sqrt(s)=8 TeV},%
pdfsubject={CMS},%
pdfkeywords={CMS, physics, supersymmetry}}

\maketitle 

\section{Introduction\label{sec:introduction}}
In the standard model (SM), proton-proton collision events
having a final state with isolated leptons
of the same sign are extremely rare.
Searches for anomalous
production of same-sign dileptons can therefore be very sensitive to new physics
processes that produce this signature copiously.
These include
supersymmetry (SUSY)~\cite{Barnett:1993ea,Guchait:1994zk,Baer:1995va},
universal extra dimensions~\cite{Cheng:2002ab},
pair production of $T_{5/3}$ particles (fermionic partners of the top
quark)~\cite{Contino:2008hi},
heavy Majorana neutrinos~\cite{Almeida:1997em},
and same-sign top-quark pair production~\cite{Han:2009,fcnczprime}.
In SUSY, for example, same-sign dileptons occur naturally with the production
of gluino pairs, when each gluino decays to a top quark and a top anti-squark, with
the anti-squark further decaying into a top anti-quark and a neutralino.

In this paper we describe searches for new physics with same-sign dileptons ($\Pe\Pe$, $\Pe\Pgm$, and $\Pgm\Pgm$)
and hadronic jets, with or without accompanying missing transverse energy (\ETmiss).
Our choice of signatures is driven by the following considerations.
New physics signals with large cross sections are likely to be produced by strong
interactions, and we thus expect significant hadronic activity in conjunction with the
two leptons. Astrophysical evidence
for dark matter~\cite{Feng:2010gw} suggests considering
SUSY models with $R$-parity conservation, which provides an excellent dark matter candidate ---
a stable lightest supersymmetric particle (LSP)
that escapes detection. Therefore, a search for this signature
involves sizable \ETmiss due to undetected LSPs.
Nevertheless, we also consider signatures without significant \ETmiss in order to be sensitive to
SUSY models with $R$-parity violation (RPV)~\cite{Martin:SusyPrimer} which imply
an unstable LSP.
Beyond these general guiding principles, the choice of signatures
is made independently of any particular physics model and, as a result, these
signatures can be applied also to probe non-supersymmetric extensions of the SM.

The results reported in this document expand upon a previous search~\cite{sspaper8TeV10fb} and are based on
the proton-proton collision dataset at $\sqrt{s}= 8$\TeV collected with the
Compact Muon Solenoid (CMS) detector at the Large Hadron Collider (LHC) during 2012,
corresponding to an integrated luminosity of 19.5\fbinv.
We consider several final states, characterized by the scalar sum (\HT) of the
transverse momenta (\pt) of jets, \ETmiss,
the number of jets, and the number of jets identified as originating from b quarks (b-tagged jets).
Additionally, in order to provide coverage for a wide range of generic signatures,
we perform the analysis with two different requirements on the lepton \pt:
the high-\pt~analysis, where the leptons are selected
with a \pt~requirement of at least 20\GeV, and the low-\pt~analysis, where the \pt~threshold is
lowered to 10\GeV. While the low-\pt leptons extend the sensitivity to scenarios
with a compressed spectrum of SUSY particle masses, the high-\pt~analysis targets models where
the leptons are produced via on-shell $\PW$ or $\Z$ bosons, and is less subject to backgrounds
with leptons originating from jets.
The use of a lower threshold on lepton \pt for the low-\pt analysis is compensated
by a tighter \HT requirement. In this respect, the two searches are complementary,
even if partially overlapping.

In contrast to the previous analysis~\cite{sspaper8TeV10fb}, the signal regions within each of the low-
and high-\pt analyses are defined to be exclusive.  Furthermore, we increase the number of search regions in order
to improve the sensitivity to a wider class of beyond-standard-model (BSM) processes.
The selection criteria for the analysis objects and the methods used to estimate the SM backgrounds are
largely unchanged from those of our previous same-sign dilepton studies~\cite{sspaper2010,sspaper2011, Chatrchyan:2012sa,sspaper8TeV10fb}.

Tables of observed yields and estimated SM backgrounds are provided for both
the high-\pt and low-\pt analyses in each exclusive signal region.
Having found no evidence for a BSM contribution to the event counts, limits are set on a variety of
SUSY-inspired models by performing a counting experiment in each exclusive search region.
Additionally, results for the high-\pt~analysis are used to set upper limits on the cross sections of
the same-sign top-quark pair production and quadruple top-quark production,
which can arise from new physics or as rare processes in the SM.

Finally, we include additional information on the event selection efficiencies to facilitate the
interpretation of these results within models not considered in this paper.

\section{The CMS detector\label{sec:detector}}
The central feature of the CMS apparatus is a superconducting solenoid,
of 6\unit{m} internal diameter, providing a magnetic
field of 3.8\unit{T}.
The experiment uses a right-handed coordinate system, with the origin defined to be the nominal interaction point,
the $x$ axis pointing to the centre of the LHC ring,
the $y$ axis pointing up,
and the $z$ axis pointing in the anticlockwise-beam direction.
The polar angle $\theta$ is measured from the positive $z$ axis and the azimuthal angle $\phi$
is measured in the $x$-$y$ (transverse) plane.
The pseudorapidity $\eta$ is defined as $\eta = - \ln{[\tan{(\theta/2)}]}$.
Within the magnetic field volume are a silicon pixel and strip tracker, a crystal electromagnetic calorimeter,
and a brass-scintillator hadron calorimeter.
Muons are measured in gas-ionization detectors embedded in the steel flux-return yoke. 
Full coverage is provided by the tracker, calorimeters, and muon detectors within $\abs{\eta}< 2.4$.
In addition to the barrel and endcap calorimeters up to $\abs{\eta}=3$,
CMS has extensive forward calorimetry reaching $\abs{\eta} \lesssim 5$.
Events are selected by a two-stage trigger system: a hardware-based trigger (L1)
followed by a software-based high-level trigger (HLT) running on the data acquisition computer farm.
A more detailed description of the
CMS apparatus can be found in Ref.~\cite{JINST}.

\section{Event selection and Monte Carlo simulation}
\label{sec:evsel}
Events used in this search are selected using two complementary online algorithms.
The high-\pt analysis uses a set of dilepton triggers,
requiring the first (second) highest-\pt lepton to have $\pt > 17$~(8)\GeV at the HLT.
The low-\pt analysis uses
high-level triggers that employ a reduced \pt threshold on leptons, of 8\GeV, and looser
lepton identification requirements, but apply an additional online selection
of $\HT>175$\GeV. The minimum lepton \pt, the lepton identification requirements, and
the \HT selections that are imposed offline for these two analyses are driven by the trigger selections.
The selection efficiencies of these triggers for events used in this analysis vary between
81\% and 96\% and are discussed in detail in Section~\ref{sec:eff}.

Offline, events with at least two isolated same-sign leptons ($\Pe\Pe$, $\Pe\Pgm$ or $\Pgm\Pgm$)
and at least two jets  are selected.
The lepton pairs are required to have an invariant mass above 8\GeV and to be consistent with originating
from the same collision vertex.
The requirement on the transverse impact parameter, calculated
with respect to the primary vertex, has been tightened to 100~(50)~\mum for electrons (muons)
compared to the previous versions of this analysis.
This selection further suppresses the backgrounds from two sources: non-prompt leptons from
semi-leptonic decays of heavy-flavour quarks and lepton charge misidentification.
The algorithms used to calculate the isolation of the leptons, reconstruct jets, identify b-tagged jets,
as well as the jet-lepton separation requirements are identical to the ones described in Refs.~\cite{sspaper2010, sspaper8TeV10fb}.
For the identification of b-quark jets we continue to use the medium operating point of the combined secondary vertex (CSV)
algorithm~\cite{beff:2012}, which is based on the combination of secondary-vertex reconstruction
and track-based lifetime information.
The treatment of the effects of multiple proton-proton
interactions within the same LHC bunch-crossing (pileup) on jet energies~\cite{Cacciari:2007fd} also remains unchanged.
Unlike the previous analysis, there is no requirement on the number of b-tagged jets when selecting events.
This number is, however, used in the categorization of events into various signal regions.

\begin{table}[h]
\begin{center}
\topcaption{\label{tab:kin} Kinematic and fiducial requirements on leptons and jets that are used to define the low-\pt (high-\pt)
analysis.}
\begin{tabular}{l|c|c}
\hline \hline
Object              & $\pt$ (\GeVns{})         & $\abs{\eta}$ \\ \hline
Electrons     & ${>}10 (20)  $ & ${<}2.4$ and ${\notin}[1.4442,1.566]$ \\
Muons         & ${>}10 (20)  $ & ${<}2.4$ \\
Jets          & ${>}40$ & ${<}2.4$ \\
b-tagged jets & ${>}40$ & ${<}2.4$ \\
\hline \hline
\end{tabular}
\end{center}
\end{table}
Kinematic selections for jets, leptons, and b-tagged jets are summarized in Table~\ref{tab:kin}.
Events with a third lepton are rejected if the lepton forms an
opposite-sign same-flavour pair with one of the first
two leptons for which the invariant mass of the pair ($m_{\ell \ell}$) satisfies
$m_{\ell \ell} < 12$\GeV ($\pt > 5$\GeV)
or $76 < m_{\ell \ell} < 106$\GeV ($\pt > 10$\GeV).
These requirements are designed to minimize backgrounds from
processes with a low-mass bound state or $\gamma^* \to \ell^+ \ell^-$ in the
final state, as well as multiboson ($\PW\cPZ$, $\cPZ\cPZ$, and triboson) production.

Monte Carlo (MC) simulations, which include pileup effects, are used to estimate some of the SM backgrounds
(see Section~\ref{sec:bg}), as well as to calculate the efficiency for
various new physics scenarios.
All SM background samples are generated with the
\MADGRAPH~5~\cite{MADGRAPH5} program and simulated using a \GEANTfour-based model~\cite{Geant}
of the CMS detector. Signal samples are produced with \MADGRAPH~5 using the CTEQ6L1~\cite{cteq}
parton distribution functions (PDF); up to two additional partons are
present in the matrix element calculations. Version~6.424 of \PYTHIA~\cite{PYTHIA}
is used to simulate parton showering and hadronization, as well as the decay of SUSY particles.
A signal sample for an RPV model is produced with \PYTHIA~6.424.
For signal samples, the detector simulation is performed using the CMS fast
simulation package~\cite{FastSim}. Detailed cross checks are performed
to ensure that the results obtained with fast simulation are in
agreement with the ones obtained with $\GEANT$-based detector simulation.
Simulated events are processed with the same chain of reconstruction
programs that is used for data.

\section{Search strategy}
\label{sec:searchStrategy}

The search is based on comparing the number of observed events with the expectation
from SM processes in several signal regions (SR) that have different requirements on
four discriminating variables:
\ETmiss, \HT, the number of jets, and the number of b-tagged jets.
We define two sets of signal regions: \emph{baseline} and \emph{final} SRs.
The former set imposes looser selection requirements, thereby forming a sample of events
where the contributions of signal events are expected to be negligible, that
is used to validate methods that are employed
to predict the background in the final SRs; the latter set is based on tighter
selection requirements, making it sensitive to
many BSM processes.
The interpretation of the results, discussed in Section~\ref{sec:models}, is primarily based on the final SRs.

Search regions defined in bins of the number of jets and b-tagged
jets provide broad coverage of strongly produced SUSY particles, including
signatures with low hadronic activity as well as signatures involving third-generation
squarks.
Additionally, as SUSY models with a small mass splitting between the parent sparticle and the LSP
may result in low~\ETmiss, we also define search regions with a looser requirement on~\ETmiss.
The high-\pt search is
ideal for BSM models with an on-shell \PW~boson produced in a new-physics particle decay,
but events with an off-shell \PW~boson can produce low-\pt leptons,
which is why leptons with transverse momenta as low as 10\GeV  are included
in this study.

\begin{table}[h]
\begin{center}
\topcaption{Definition of the baseline signal regions for the three different requirements
  on the number of b-tagged jets (\nbjets). \njets refers to the number of jets in the event.
  The same naming scheme is used for both the low- and high-\pt analyses, which
  differ only in a looser requirement on \HT (in parentheses) for the
  high-\pt analysis.}
\label{tab:baselineSR}
\begin{tabular}{c|c|c|c|c}
  \hline \hline
  \HT (\GeVns{})     & \ETmiss (\GeVns{})                   & \njets   & \nbjets & SR name\\ \hline
  $>$250 (80)        & $>$30 if $\HT < 500$ else $>$0    & $\ge$2        & $ = $0 & BSR0 \\
  $>$250 (80)        & $>$30 if $\HT < 500$ else $>$0    & $\ge$2        & $ = $1 & BSR1 \\
  $>$250 (80)        & $>$30 if $\HT < 500$ else $>$0    & $\ge$2        & $\ge$2 & BSR2 \\
  \hline \hline
\end{tabular}
\end{center}
\end{table}

We define the three baseline signal regions (BSR0, BSR1, and BSR2) for both the low- and
high-\pt analyses, as described in Table~\ref{tab:baselineSR}.
The event selection criteria are tightened and the granularity of the regions is increased to
define the 24 final SRs described in Table~\ref{tab:srs} for the high-\pt analysis.
For the low-\pt signal regions, the categories are equivalent to those of the high-\pt
analysis, but the selection differs in the requirement on \HT and lepton \pt.
The threshold on \HT is increased from 200 to 250\GeV in order to ensure 100\% efficiency
for the triggers used by the low-\pt event selection.
All 24 signal regions are mutually exclusive and may therefore be statistically combined within either
high-\pt or low-\pt analysis.

\begin{table}[h]
	\begin{center}
	\topcaption{Definition of the signal regions for the high-\pt analysis. The low-\pt analysis employs a tighter
	requirement $\HT > 250$\GeV and uses the same numbering scheme, in which the first digit in the name
	represents the requirement on the number of b-tagged jets for that search region, e.g. SR01, SR11, and SR21
	correspond to SRs with \nbjets 0, 1, and $\ge$2, respectively.
	  }
	\label{tab:srs}
		\begin{tabular}{c|c|c|c|c}
		\hline \hline
		         \nbjets           & \ETmiss  (\GeVns{})         & \njets   & \HT $\in [200,400]$~(\GeVns{}) & $\HT >400$ (\GeVns{}) \\ \hline
		\multirow{4}{*}{$=0$}  & \multirow{2}{*}{50--120} & 2--3      &    SR01      &    SR02     \\ 
		                           &                         & ${\geq}4$ &    SR03      &    SR04     \\ \cline{2-5}
		                           & \multirow{2}{*}{${>}120$} & 2--3      &    SR05      &    SR06     \\ 
		                           &                         & ${\geq}4$ &    SR07      &    SR08     \\ \hline
		\multirow{4}{*}{$=1$}      & \multirow{2}{*}{50--120} & 2--3      &    SR11      &    SR12     \\ 
		                           &                         & ${\geq}4$ &    SR13      &    SR14     \\ \cline{2-5}
		                           & \multirow{2}{*}{${>}120$} & 2--3      &    SR15      &    SR16     \\ 
		                           &                         & ${\geq}4$ &    SR17      &    SR18     \\ \hline
		\multirow{4}{*}{$\geq 2$}  & \multirow{2}{*}{50--120} & 2--3      &    SR21      &    SR22     \\ 
		                           &                         & ${\geq}4$ &    SR23      &    SR24     \\ \cline{2-5}
		                           & \multirow{2}{*}{${>}120$} & 2--3      &    SR25      &    SR26     \\ 
		                           &                         & ${\geq}4$ &    SR27      &    SR28     \\
                \hline \hline
		\end{tabular}
	\end{center}
\end{table}

Additional (overlapping) signal regions, listed in Table~\ref{tab:srs3x}, are defined with no or loose \ETmiss requirements
in order to provide better sensitivity
to scenarios such as RPV SUSY models
and same-sign top-quark pair production. These search regions are formed using events that satisfy
high-\pt lepton selection and contain at least two jets.
Because in RPV SUSY scenarios the LSP decays, mainly into detectable leptons and quarks, such events are not expected to have large \ETmiss, but they usually have  substantial \HT. Thus, in search regions designed for such models, the \ETmiss requirement is removed completely, while a relatively high $\HT > 500\GeV$ requirement is applied to reduce the level of SM background. These search regions are labelled as RPV0 and RPV2 for $\nbjets \ge0$  and ${\ge}2$, respectively.

Same-sign top quark pair events in which the \PW~bosons decay leptonically generally contain
moderate \ETmiss, due to the accompanying neutrinos.
Using events with $\ETmiss >30$\GeV, we form four signal regions,
denoted SStop1, SStop2, SStop1++, and SStop2++, where "++" refers to the selection of only positively
charged dilepton pairs. Note that in most new physics scenarios,
$\Pp \Pp \to \cPaqt \cPaqt$ is suppressed with respect to $\Pp \Pp \to \cPqt \cPqt$
because the PDF of the proton is dominated by quarks, rather than anti-quarks.
For such scenarios, the SStop1++
and SStop2++ signal regions are expected to provide higher sensitivity.

\begin{table}[h]
	\begin{center}
	\topcaption{Signal regions that are used in the search for same-sign top-quark pair production and RPV SUSY processes.}
	\label{tab:srs3x}
		\begin{tabular}{c|c|c|c|c|l}
		\hline \hline
	 \njets   &	\nbjets    & \ETmiss (\GeVns{})    & \HT (\GeVns{})   & Lepton charge  & SR name    \\ \hline
	 $\geq$2 &	 $\geq$0  &  ${>}0$      & ${>}500$ & ++/-- --   &  RPV0		   \\
	 $\geq$2 &	 $\geq$2  &  ${>}0$      & ${>}500$ & ++/-- --   &  RPV2		   \\
	 $\geq$2 &	 $=$1       &  ${>}30$     & ${>}80$  & ++/-- --   &  SStop1	  \\
	 $\geq$2 &	 $=$1       &  ${>}30$     & ${>}80$  & ++ only    &  SStop1++        \\
	 $\geq$2 &	$\geq$2   &  ${>}30$     & ${>}80$  & ++/-- --   &  SStop2          \\
	 $\geq$2 &	$\geq$2   &  ${>}30$     & ${>}80$  & ++ only    &  SStop2++        \\
         \hline \hline
		\end{tabular}
	\end{center}
\end{table}

\section{Backgrounds}
\label{sec:bg}

There are three main sources of SM background in this analysis, which are described below. 
More details on the methods used to estimate these backgrounds can be found in
Refs.~\cite{Chatrchyan:2012sa, sspaper2010}.

\begin{itemize}

\item ``Non-Prompt leptons'', i.e. leptons from heavy-flavour decays,
misidentified hadrons, muons from light-meson decays in flight, or electrons
from unidentified photon conversions.  The background caused by these non-prompt
leptons, which is dominated by \ttbar and \wjets processes, is estimated from a
sample of events with at least one lepton that passes a loose
selection but fails the full set of tight identification
and isolation requirements described in Section~\ref{sec:evsel}. 
The background rate is obtained 
by scaling the number of events in this sample by a
``tight-to-loose'' ratio, i.e. the probability that
a loosely identified non-prompt lepton also passes the full
set of requirements. Various definitions of the loose lepton selection criteria are studied in detail, and 
combination of relaxed isolation and lepton-identification requirements is used.
These probabilities are measured as a function of lepton \pt and $\eta$, as well as event kinematics, 
in control samples of QCD multijet events that are enriched in non-prompt leptons.

\item Rare SM processes that yield same-sign leptons, mostly from 
$\ttbar\PW$, $\ttbar\cPZ$, and diboson production.
We also include the contribution 
from the SM Higgs boson produced in association with a vector boson or a pair of top quarks 
in this category of background.
All these backgrounds are estimated from MC simulation. The event 
yields are corrected for several effects, summarized in Section~\ref{sec:eff}, 
to account for the differences between object selection efficiencies in data and simulation.

\item Charge misidentification, i.e. events with opposite-sign isolated
leptons where the charge of one of the leptons is 
misidentified because of severe bremsstrahlung in the tracker material.
This background, which is relevant only for electrons and is negligible for muons,
 is estimated by selecting opposite-sign \Pe \Pe\
or \Pe \Pgm\
events passing the full kinematic selection and then weighting them by the
\pt- and $\eta$-dependent probability of electron charge
misassignment.  This probability, which varies between $10^{-4}\text{ and } 10^{-5}$, is 
obtained from simulation and is then validated with a control data sample of $\cPZ \to \Pe \Pe$ events.

\end{itemize}

Backgrounds stemming from non-prompt leptons constitute the major contribution 
to the total background in most search regions. The rare SM processes dominate in 
the search regions with large numbers of b-tagged jets or high $\ETmiss$ requirements. 
The contribution from charge misidentification is generally much smaller and stays 
below the few-percent level in all search regions.

The primary origin of the systematic uncertainty for the 
non-prompt lepton background estimate is differences
between the QCD multijet sample, where the ``tight-to-loose'' 
ratio is determined, and the signal regions, where the method 
is applied, both for the event kinematics and for the relative 
rates of the various sources of non-prompt leptons.
A systematic uncertainty also arises because \ttbar and \wjets events, the two dominant 
components of the non-prompt background, differ themselves in the event kinematics 
and relative importance of the various sources, making it difficult to define a ``tight-to-loose'' 
ratio that is equally appropriate for both components.
Based on the variation between true and predicted background 
yields when the background estimation method is applied to simulation, the systematic 
uncertainty of the estimate is assessed at 50\%.
This systematic part is the dominant uncertainty in the non-prompt lepton background 
estimate in most signal regions.  The statistical uncertainty in the method is driven by 
the number of events in the sideband regions, defined with relaxed lepton requirements, that are 
used to estimate the non-prompt lepton background.  As the kinematic selections are tightened, 
the statistical uncertainty becomes more important, becoming comparable in size to 
the systematic uncertainty in the search regions with the tightest selections.

For the rare SM processes, the next-to-leading-order (NLO) production 
cross sections are used to normalize the MC predictions. The cross section values used for the most relevant processes, $\ttbar\PW$ and
$\ttbar\cPZ$, are 232\unit{fb}~\cite{Campbell:2012dh}
and 208\unit{fb}~\cite{ttzNLO,Garzelli:2011is}, respectively. Because these and other rare processes are simulated using leading 
order (LO) generators, the systematic uncertainty for the rare SM background accounts both for the theoretical uncertainty 
in the cross sections and for the non-uniformity of the ratio between the LO and NLO cross sections
as a function of jet multiplicity, $\HT$, and $\ETmiss$~\cite{Campbell:2012dh}. 
The systematic uncertainties for each SM process that 
contributes to this background are assigned to be 50\% and are 
considered to be 100\% correlated across all signal  regions.

The uncertainty associated with the charge-misidentification background estimate, which is estimated to be 30\%, accounts for 
differences between data and simulation, and the limited momentum range of electrons 
probed in the control sample.

The total background in each search region is obtained by summing the yields from each of these background sources, 
and the total uncertainty is calculated by considering the individual uncertainties to be uncorrelated.

\section{Efficiencies and associated uncertainties}
\label{sec:eff}
The trigger efficiency is measured with data, using triggers that are orthogonal to those described in Section~\ref{sec:evsel}.
The measured efficiencies are summarized in Table~\ref{tab:TrigEff}.
Correction factors to take the trigger inefficiencies into account are applied to all acceptances calculated from MC simulation,
for both signal and background samples. We assign a 6\% uncertainty to these efficiencies, based on
the statistical uncertainty of the measurement and deviations from the quoted numbers
in Table~\ref{tab:TrigEff} as a function of $\abs{\eta}$  and \pt.

\begin{table}[h]
\begin{center}
\topcaption{\label{tab:TrigEff} Summary of the trigger selection efficiencies for
low- and high-\pt analyses in each channel. The thresholds on $\abs{\eta}$  and \pt
correspond to the lower \pt lepton of the dilepton pair.}
\begin{tabular}{l|c|c}
\hline \hline
Channel                            & Low-\pt         & High-\pt         \\ \hline
$\Pe \Pe$, $\pt<30$\GeV            & $0.93 \pm 0.06$ & $0.92 \pm 0.05$  \\
$\Pe \Pe$, $\pt>30$\GeV            & $0.93 \pm 0.06$ & $0.96 \pm 0.06$  \\
\Pe \Pgm\                          & $0.93 \pm 0.06$ & $0.93 \pm 0.06$  \\
$\Pgm \Pgm$,   $\abs{\eta} < 1$         & $0.94 \pm 0.06$ & $0.90 \pm 0.05$  \\
$\Pgm \Pgm$,   $1.0 < \abs{\eta} < 2.4$ & $0.90 \pm 0.05$ & $0.81 \pm 0.05$  \\
\hline \hline
\end{tabular}
\end{center}
\end{table}

The offline lepton selection efficiencies in data and simulation are measured using \cPZ-boson events to derive
simulation-to-data correction factors.
The correction factors applied to simulation are 90 (96)\% for $\pt < 20$\GeV and 94 (98)\% for $\pt > 20$\GeV for electrons (muons).
The uncertainty of the total efficiency is 5\% (3\%)
for electrons (muons) with $\pt > 15$\GeV, increasing to 10\% (5\%) for lower transverse momentum.
An additional systematic uncertainty is assigned to account for potential mismodelling of the lepton
isolation efficiency due to varying hadronic activity in signal events.  This uncertainty is 3\%
for all leptons except muons with $\pt < 30$\GeV, for which it is 5\%.

Another source of systematic uncertainty is associated with the
jet energy scale correction. This systematic uncertainty varies between
5\% and 2\% in the \pt range 40--100\GeV for jets with $\abs{\eta} < 2.4$~\cite{JES}.
It is evaluated on a single-jet basis, and its effect
is propagated to \HT, \ETmiss, the number of jets, and the number of b-tagged jets.
The importance of these effects depends on the
signal region and the model of new physics.  In general, models with high hadronic activity
and large \ETmiss are less affected by the uncertainty in the jet energy scale.
In addition, there is a contribution to the total uncertainty arising from
limited knowledge of the resolution of the jet energy, but this effect is generally
of less importance than the contribution from the jet energy scale.

The b-tagging efficiency for b-quark jets with $\abs{\eta} < 2.4$,
measured in data using samples enriched in \ttbar and muon-jet events,
has a \pt-averaged value of 0.72.  The false positive b-tagging probability for charm-quark
jets is approximately 20\%, while for jets originating from light-flavour quarks or
gluons it is of the order of 1\%.  Correction
factors, dependent on jet flavour and kinematics, are applied to simulated jets
to account for the differences in the tagging efficiency in simulation with respect to data.  The total
uncertainty of the b-tagging efficiency is determined by simultaneously varying the
efficiencies to tag a bottom, charm, or light quark up and down by
their uncertainties~\cite{beff:2012}.
The importance of this effect depends on the signal region and the
model of new physics.  In general, models with more than two
b quarks in the final state are less affected by this uncertainty.

Additional uncertainties due to possible mismodelling of the
pileup conditions or initial-state radiation (ISR)~\cite{CMS-PAS-SUS-13-011} are
evaluated and found to be 5\% and 3--15\%, respectively.
The uncertainty of the signal acceptance due to the PDF choice
is found to be less than a few percent.
Finally, there is a 2.6\% uncertainty in the yield of events
because of the uncertainty in the luminosity
normalization~\cite{CMS-PAS-LUM-13-001}.

A summary of the systematic uncertainties associated  with  the acceptance and signal efficiency for this analysis is provided
in Table~\ref{tab:signalsys}.
While the uncertainties associated with the integrated luminosity, modelling of lepton selection, trigger
efficiency, and pileup are taken to be constant across
the parameter space of the new physics models considered in this paper,
uncertainties arising from the remaining observables are estimated for each model separately
on an event-by-event basis by varying those observables within their uncertainties.
The total uncertainty in the computed acceptance
is in the 13--25\% range.
The figures in Table~\ref{tab:signalsys} are representative values for these uncertainties
and do not characterize the results for extreme kinematic regions, such as those near the diagonal
of the parameter space of the SUSY simplified models discussed in Section~\ref{sec:models},
where the particle mass spectra are compressed.

\begin{table}[h*]
	\begin{center}
		\topcaption{Summary of representative systematic uncertainties for
                  the considered signal models.}
		\label{tab:signalsys}
		\begin{tabular}{ l r}  \hline
		Source                                             & \%      \\ \hline \hline
		Luminosity                                         & 2.6     \\
		Modelling of lepton selection (ID and isolation)    & 10      \\
		Modelling of trigger efficiency                     & 6       \\
		Pileup modelling                                    & 5       \\
		Jet energy scale                                   & 1--10   \\
		Jet energy resolution                              & 0--3    \\
		b-jet identification                               & 2--10   \\
		ISR modelling                                       & 3--15   \\ \hline
		Total                                              & 13--25  \\ \hline \hline
		\end{tabular}
	\end{center}
\end{table}

\section{Results}
\label{sec:yields}
\begin{figure}[h]
\begin{center}

\ifPAPER{
\includegraphics[width=0.49\linewidth]{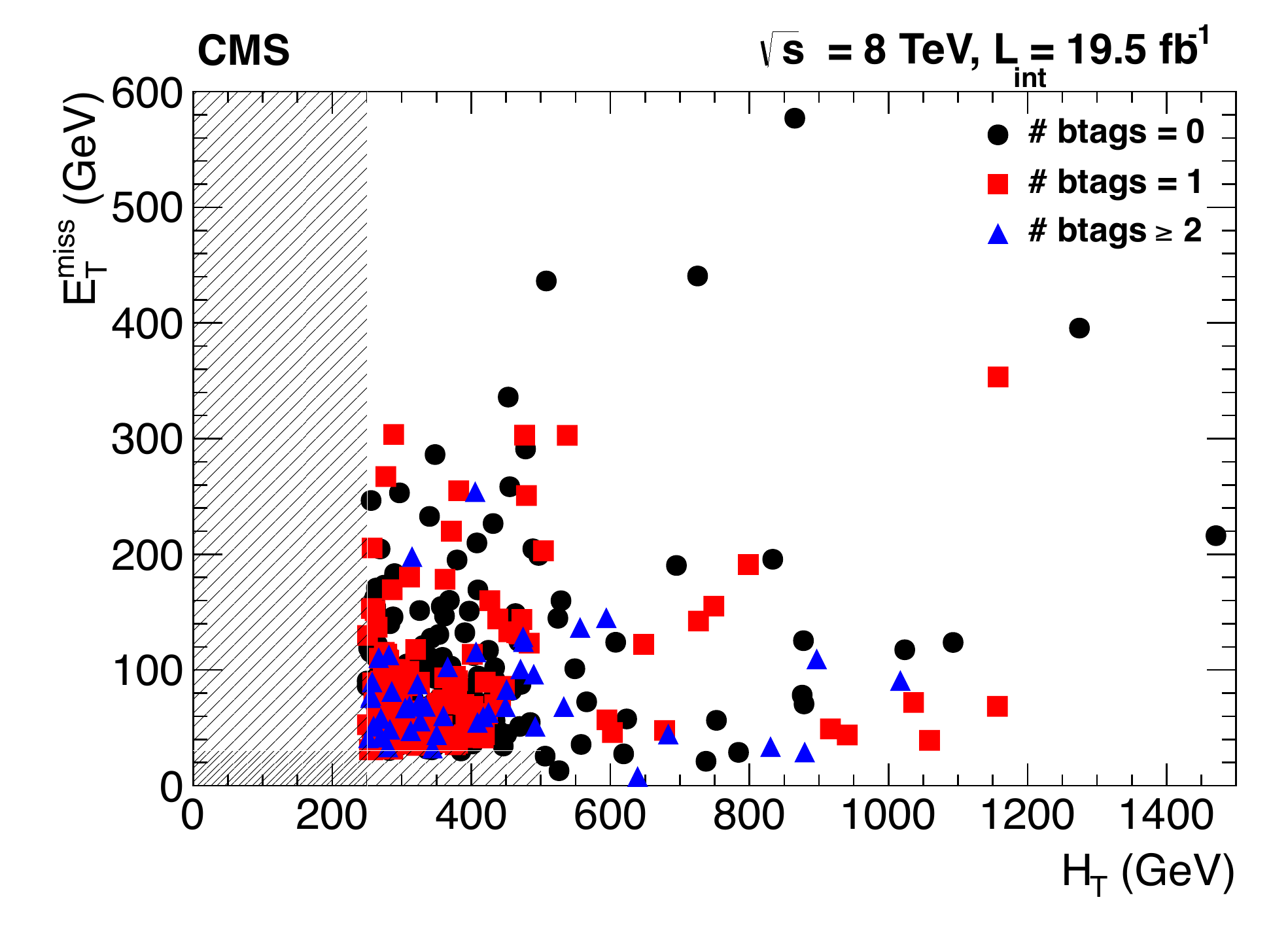} 
\includegraphics[width=0.49\linewidth]{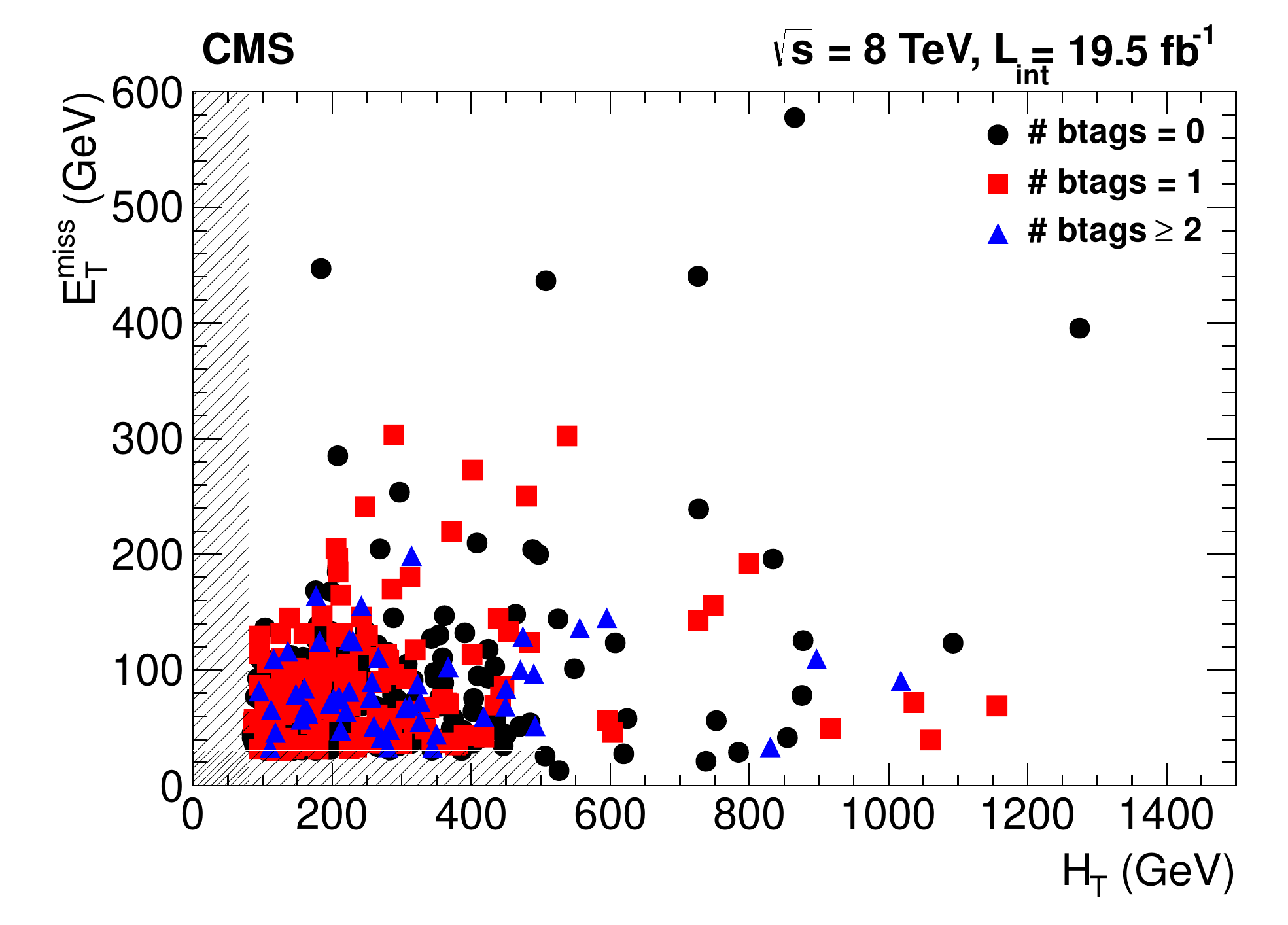}
}

\ifPAS{
\includegraphics[width=0.49\linewidth]{figs/results/PAS/HtVsMET_lpt_sr0_nbs.pdf} 
\includegraphics[width=0.49\linewidth]{figs/results/PAS/HtVsMET_hpt_sr0_nbs.pdf}
}
 
\caption{\label{fig:scatterSR0}
Distributions of \ETmiss versus \HT for the 
baseline signal regions BSR0, BSR1, and BSR2 for the low-\pt (left) and the high-\pt (right) analyses. 
The regions indicated with the hatched area are not included in the analyses.}
\end{center}
\end{figure}

\begin{figure}[!h]
\begin{center}
\ifPAPER{
\includegraphics[width=0.47\linewidth]{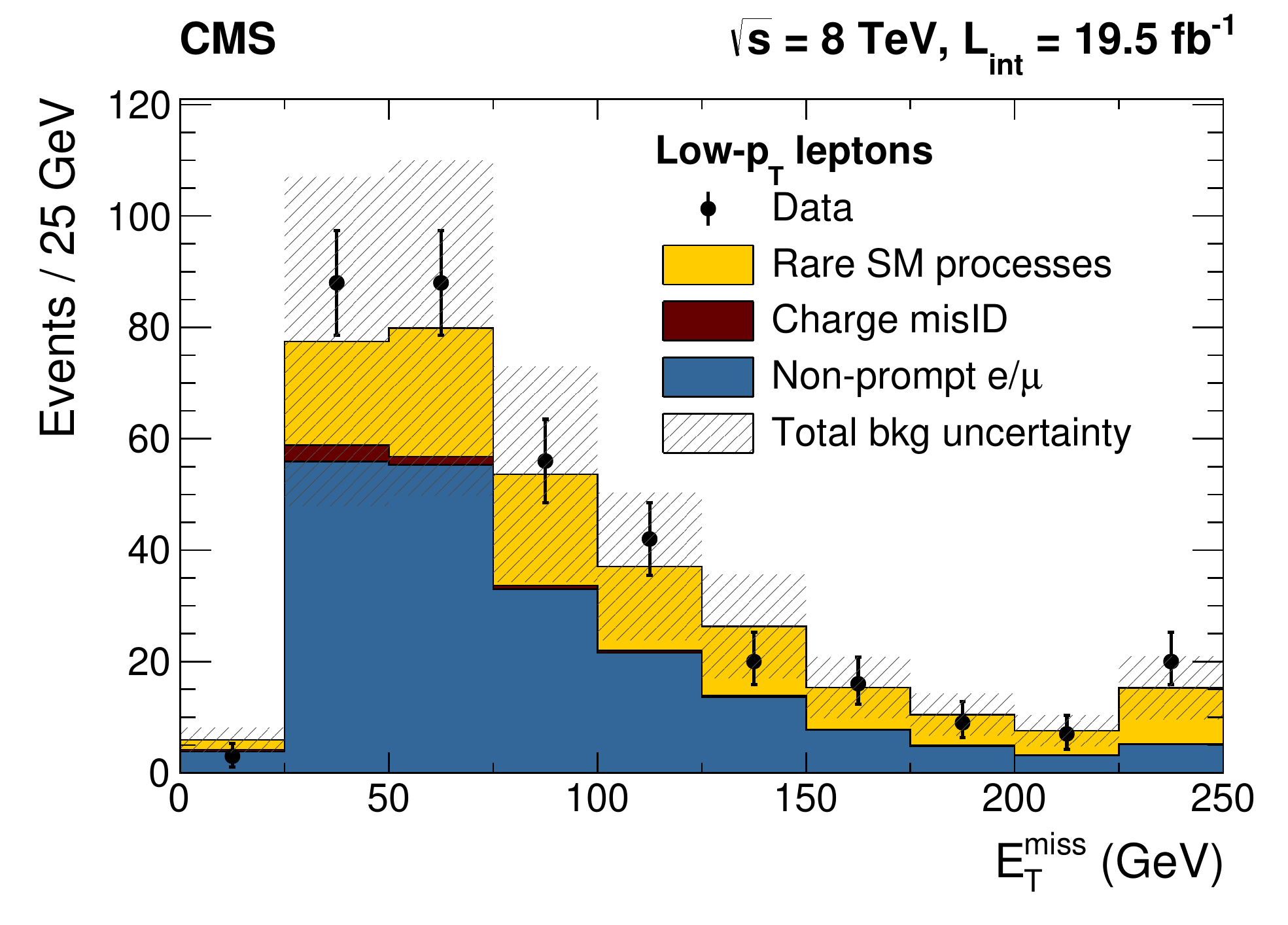} 
\includegraphics[width=0.47\linewidth]{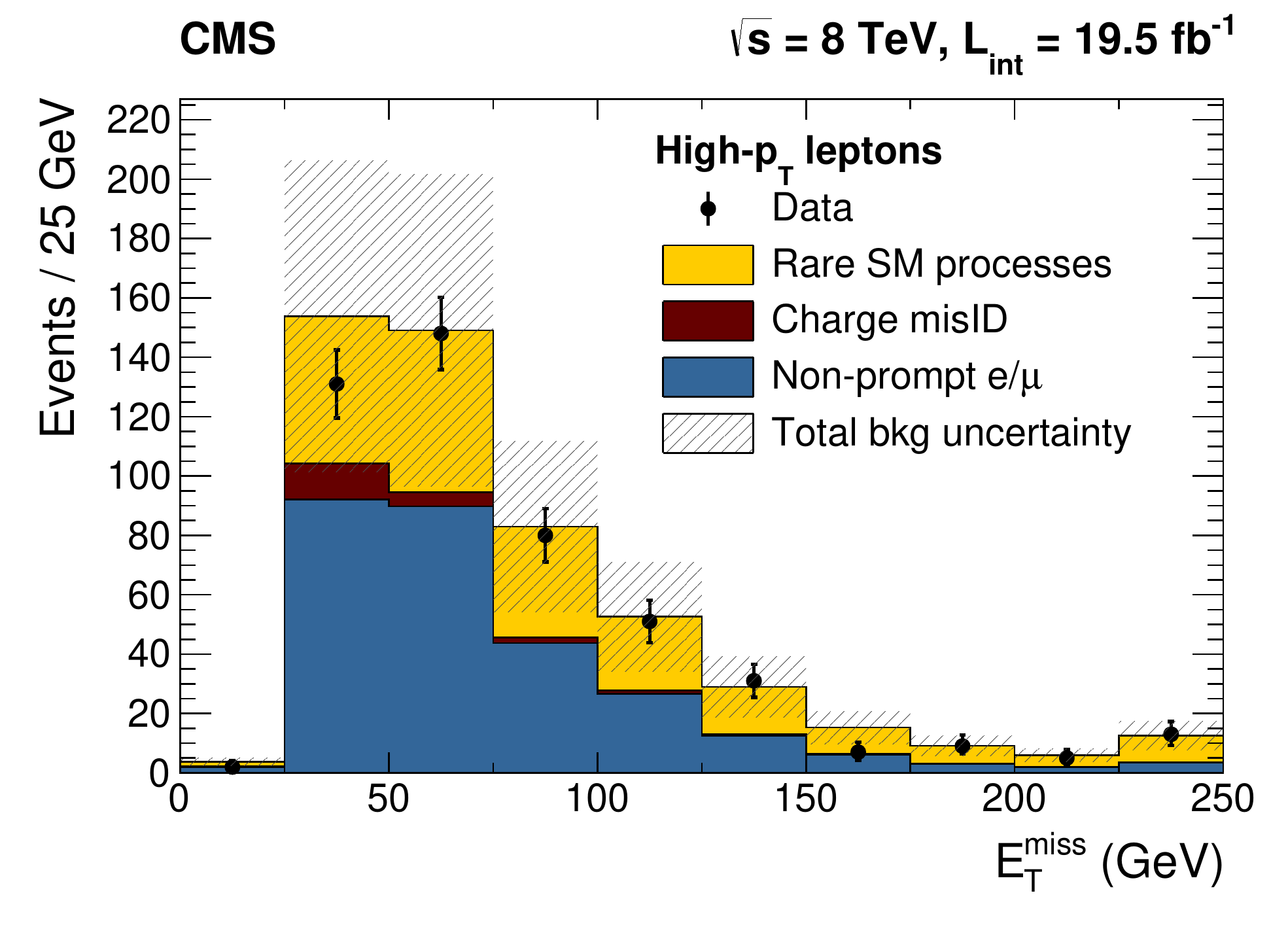} 

\includegraphics[width=0.47\linewidth]{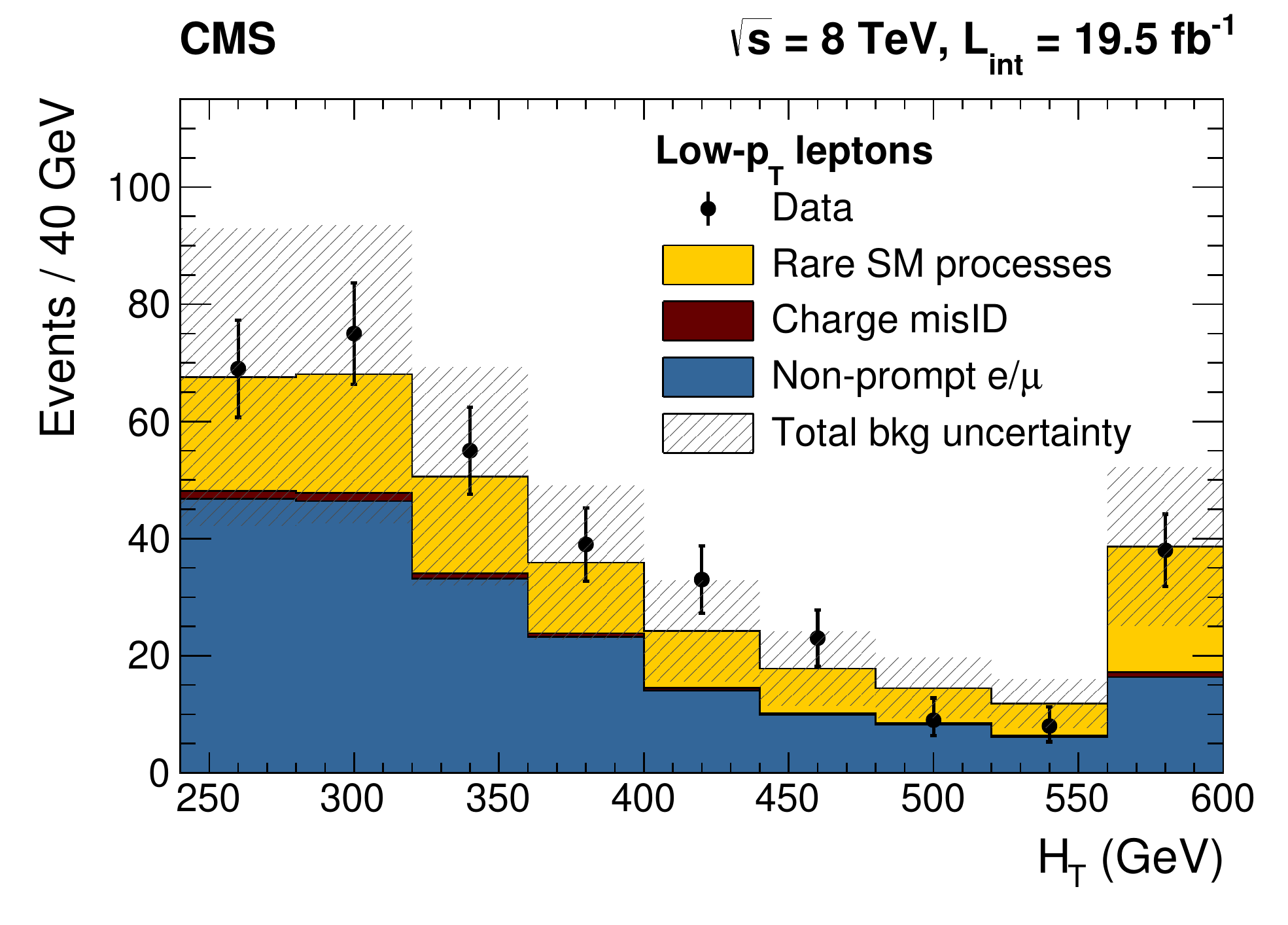} 
\includegraphics[width=0.47\linewidth]{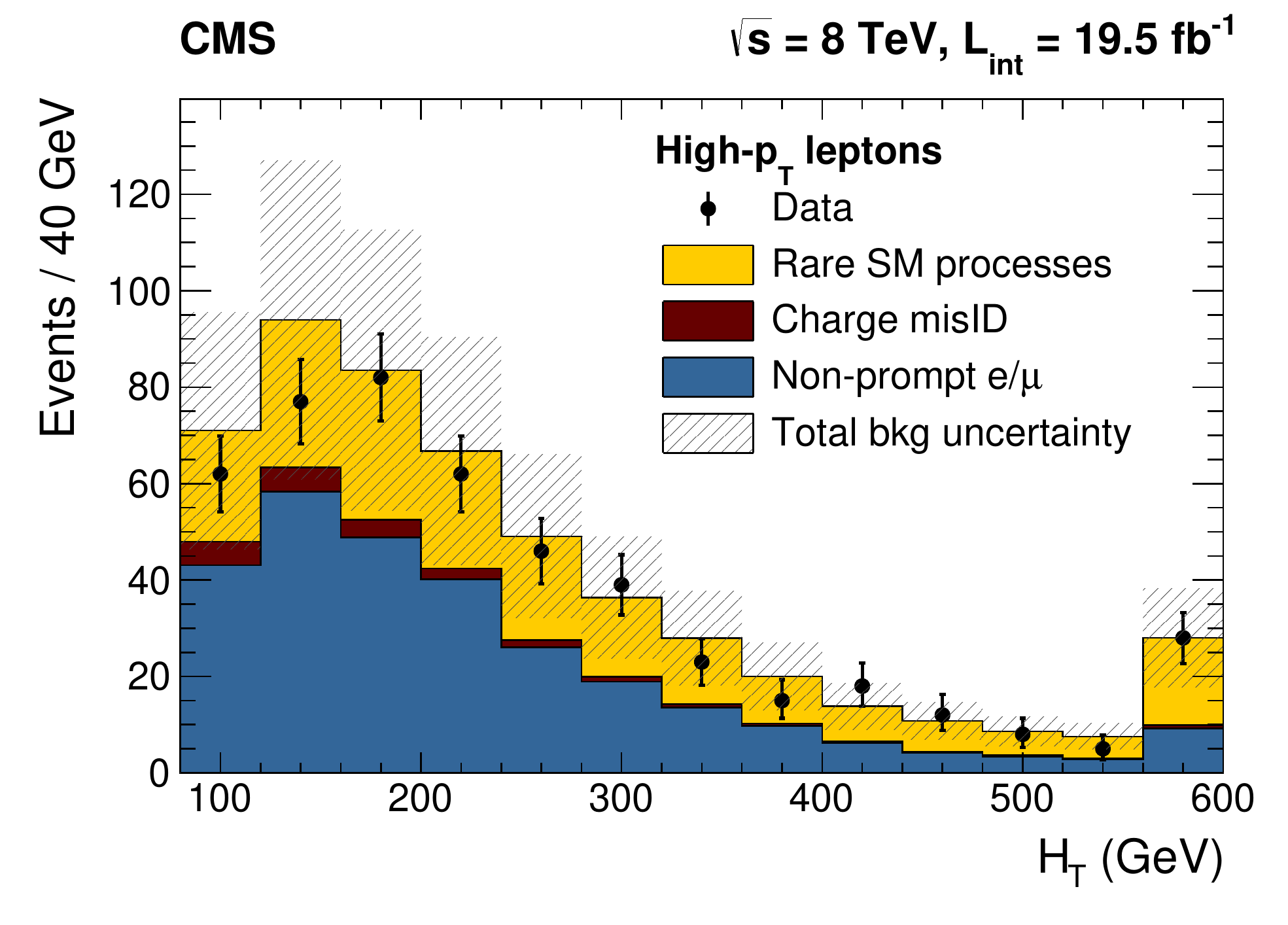} 

\includegraphics[width=0.47\linewidth]{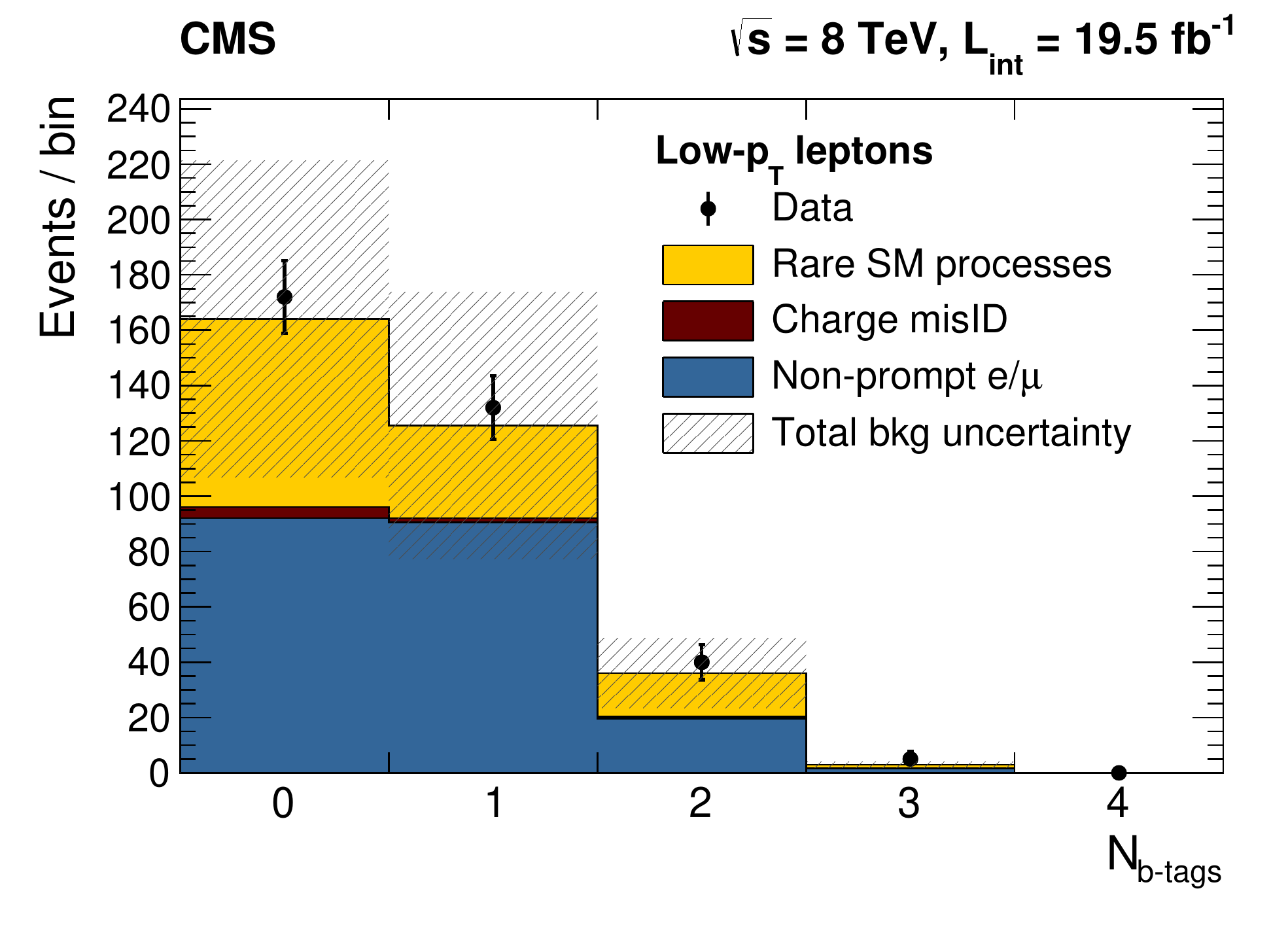} 
\includegraphics[width=0.47\linewidth]{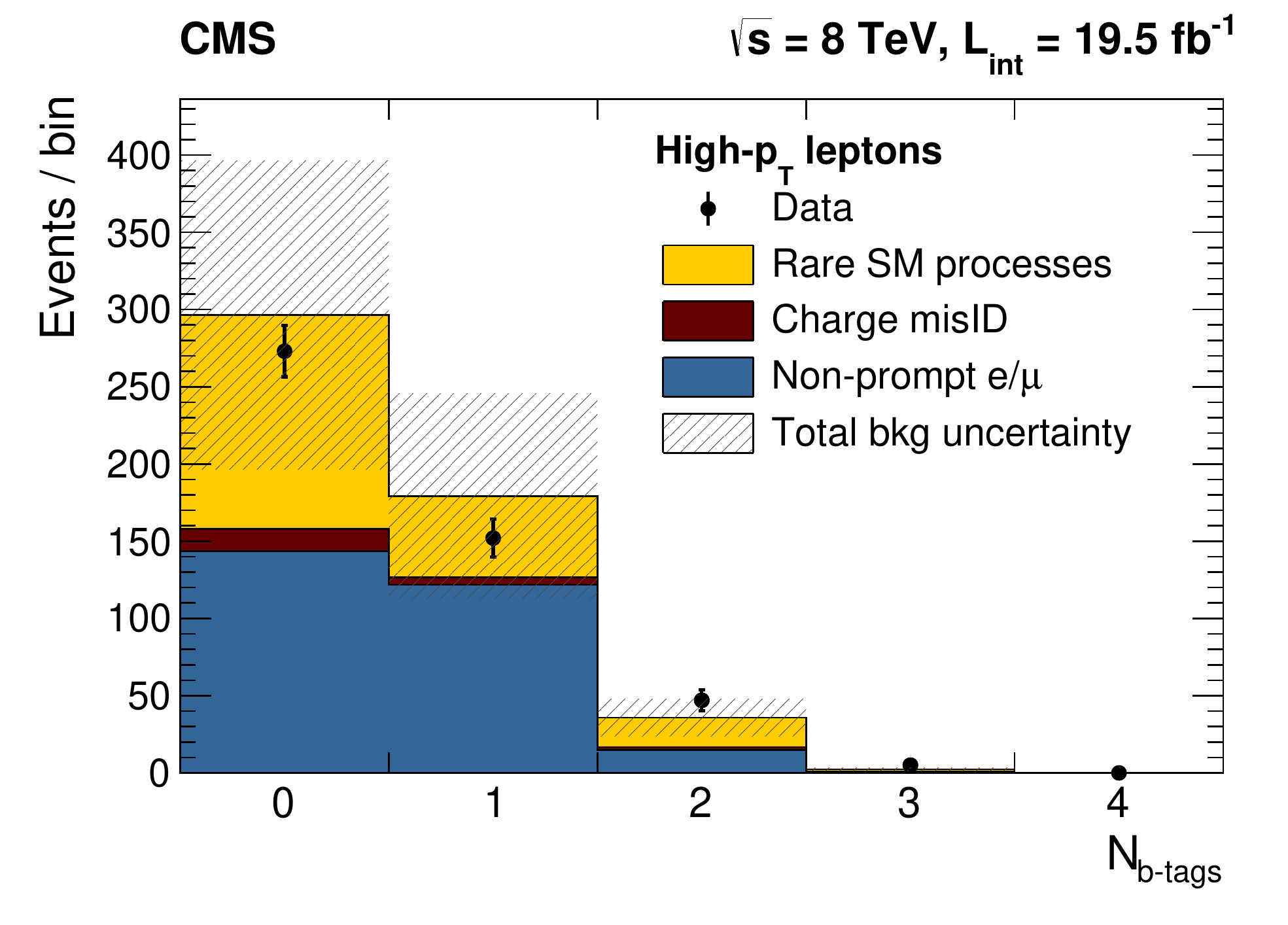} 

\includegraphics[width=0.47\linewidth]{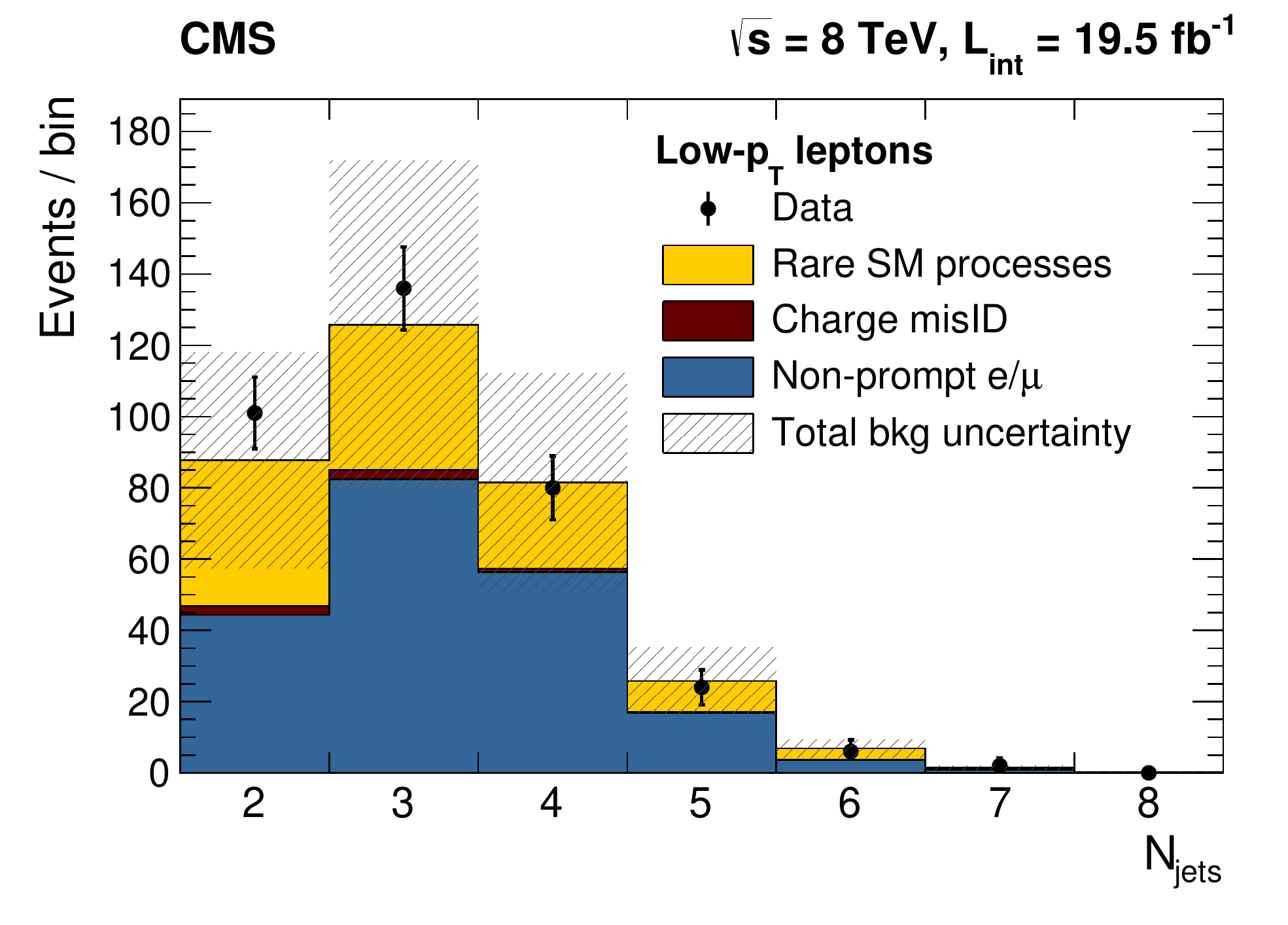} 
\includegraphics[width=0.47\linewidth]{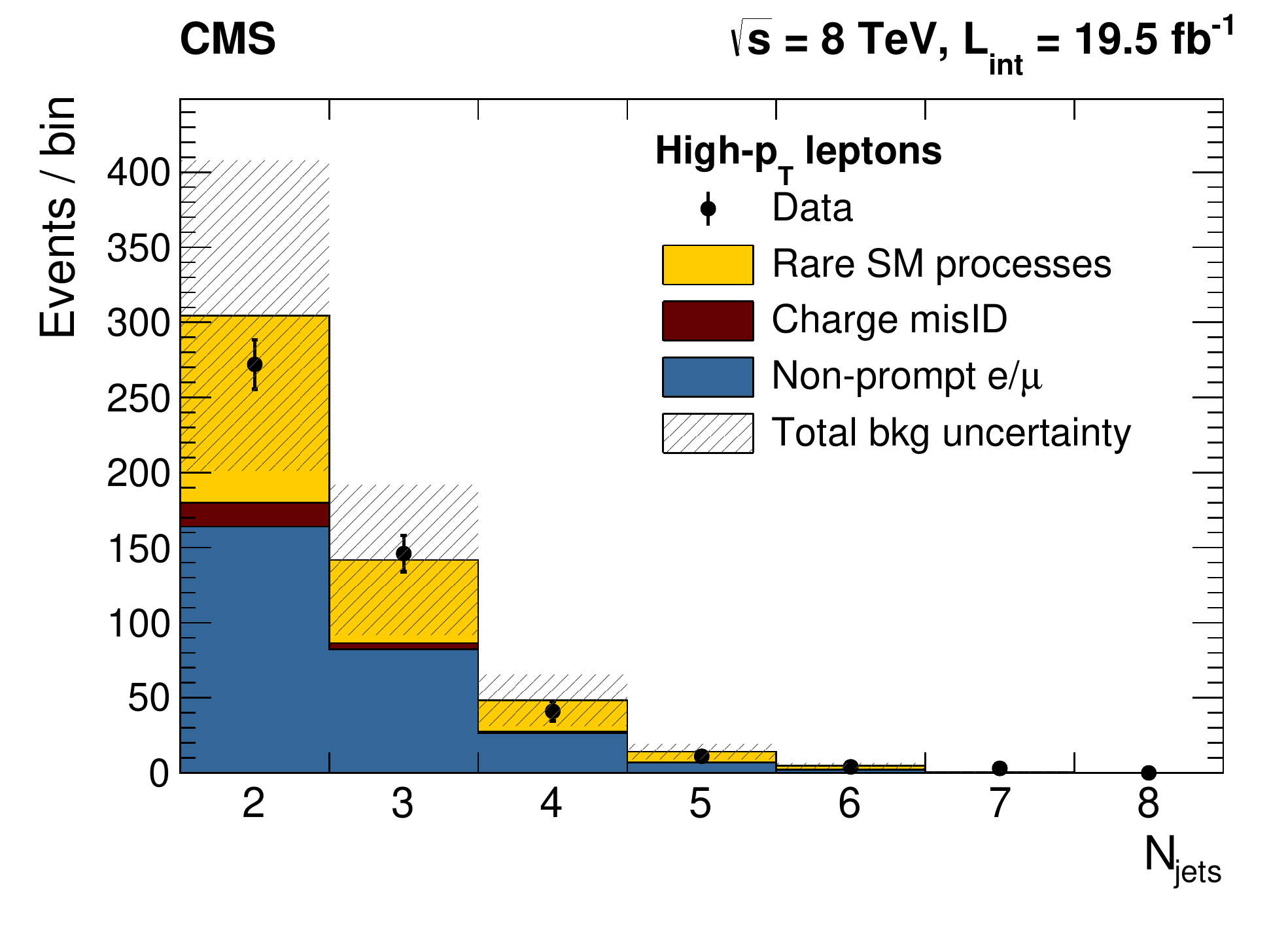} 
}

\ifPAS{
\includegraphics[width=0.47\linewidth]{figs/kinematics/PAS/kin_lowPt_histoMET.pdf} 
\includegraphics[width=0.47\linewidth]{figs/kinematics/PAS/kin_highPt_histoMET.pdf} 

\includegraphics[width=0.47\linewidth]{figs/kinematics/PAS/kin_lowPt_histoHT.pdf} 
\includegraphics[width=0.47\linewidth]{figs/kinematics/PAS/kin_highPt_histoHT.pdf} 

\includegraphics[width=0.47\linewidth]{figs/kinematics/PAS/kin_lowPt_histoNBJ.pdf} 
\includegraphics[width=0.47\linewidth]{figs/kinematics/PAS/kin_highPt_histoNBJ.pdf} 

\includegraphics[width=0.47\linewidth]{figs/kinematics/PAS/kin_lowPt_histoNJ.pdf} 
\includegraphics[width=0.47\linewidth]{figs/kinematics/PAS/kin_highPt_histoNJ.pdf} 
}
\caption{\label{fig:kin_sr0}
Distributions of \ETmiss, $\HT$, number of b-tagged jets, and number of jets for the events in the 
low-\pt (high-\pt)~baseline region with no \nbjets~requirement (events selected in BSR0, BSR1, and BSR2) 
are shown on the left (right).
Also shown as a histogram is the background prediction.
The shaded region represents the total background uncertainty.}
\end{center}
\end{figure}

The distributions of \ETmiss versus \HT for events in the three baseline signal regions 
are shown in Fig.~\ref{fig:scatterSR0}.  The results are shown separately for the low- 
and high-\pt samples.  The corresponding results for the four selection variables \HT, 
\ETmiss, \njets, and \nbjets are shown in Fig.~\ref{fig:kin_sr0}.  For these latter results, 
the SM background prediction is also shown.  There are no significant discrepancies 
observed between the observations and background predictions for any region.

\begin{figure}[!h]
	\begin{center}
\ifPAPER{
	\includegraphics[width=0.49\textwidth]{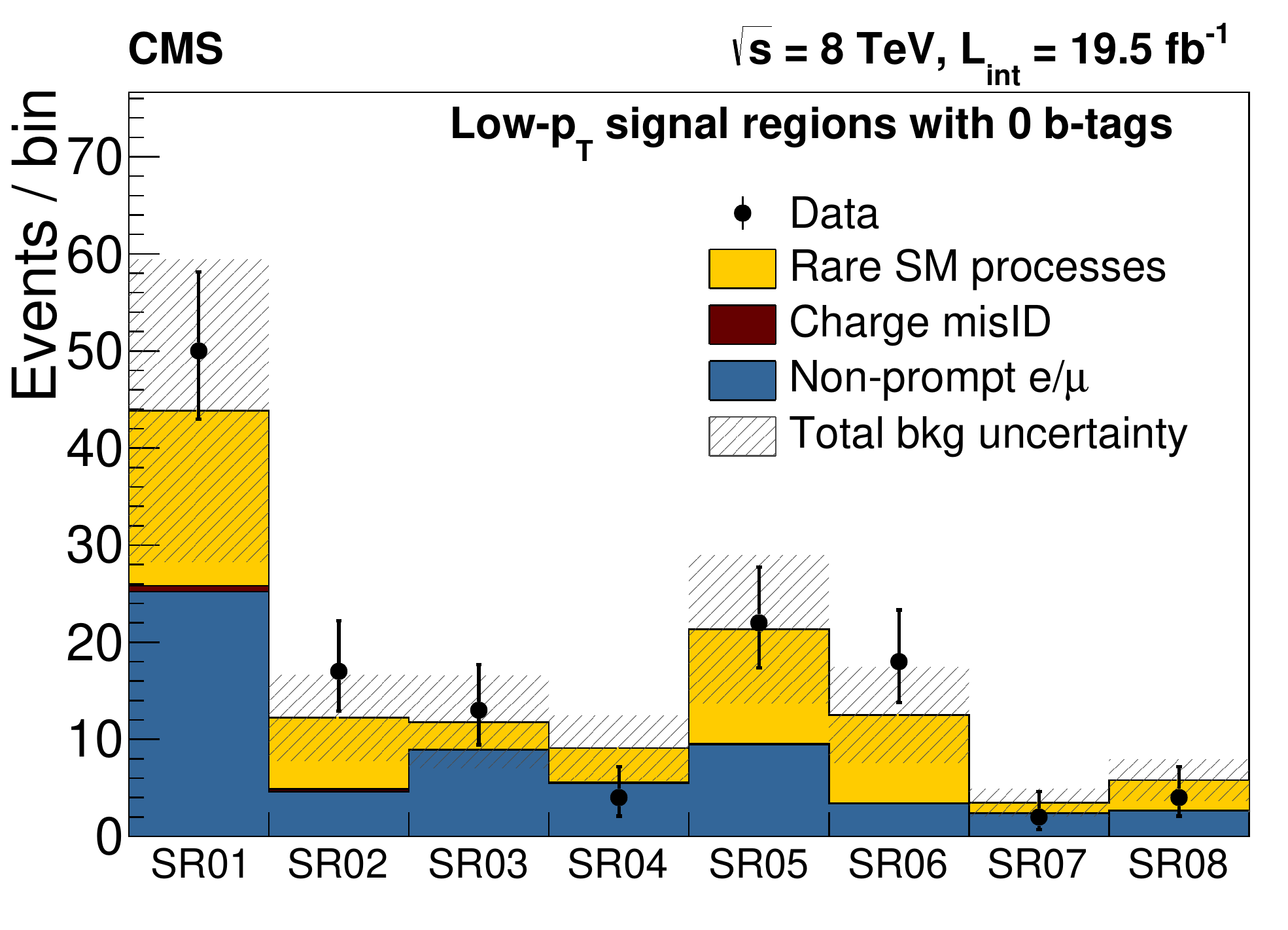} 
        \includegraphics[width=0.49\textwidth]{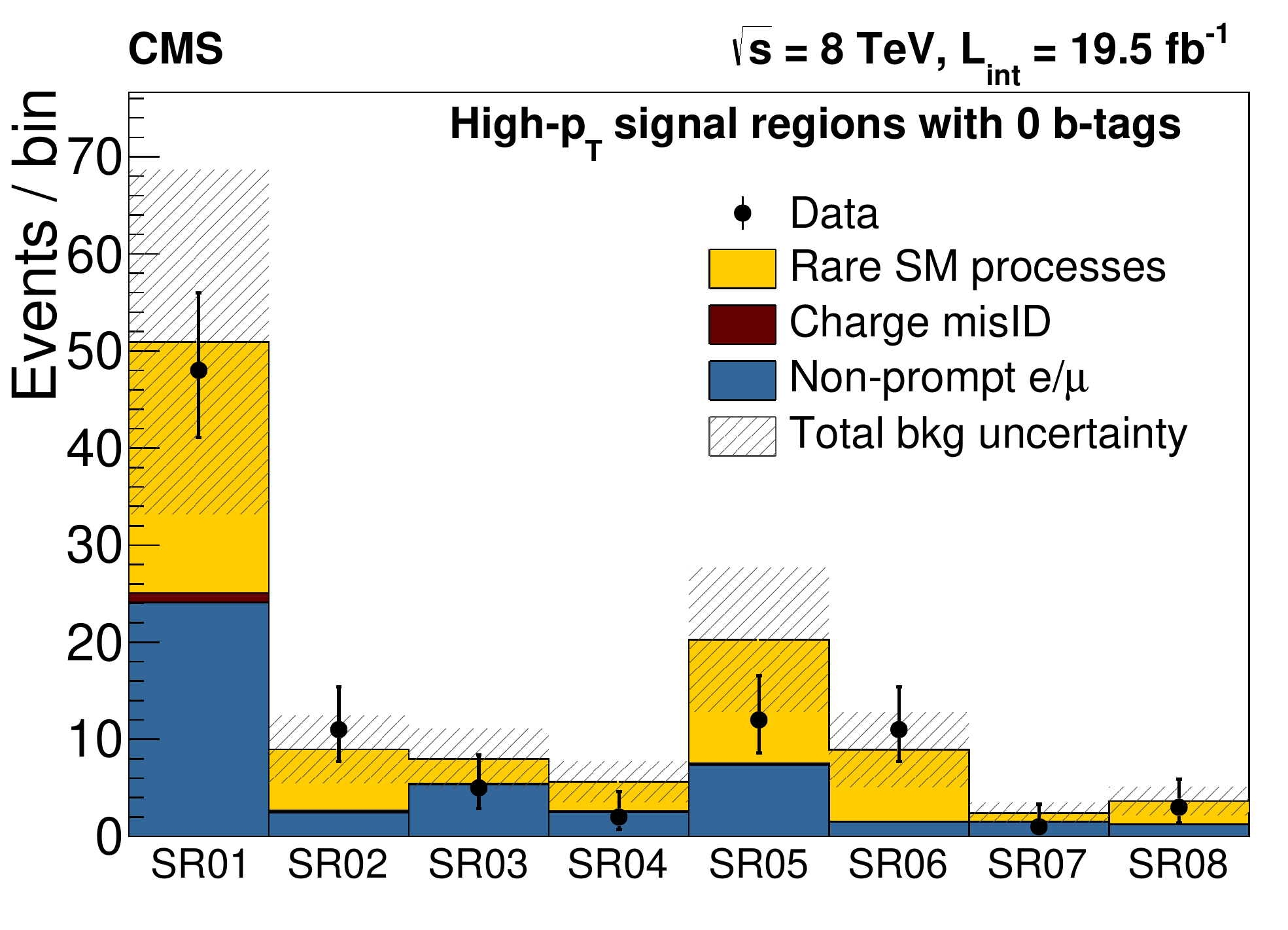}

	\includegraphics[width=0.49\textwidth]{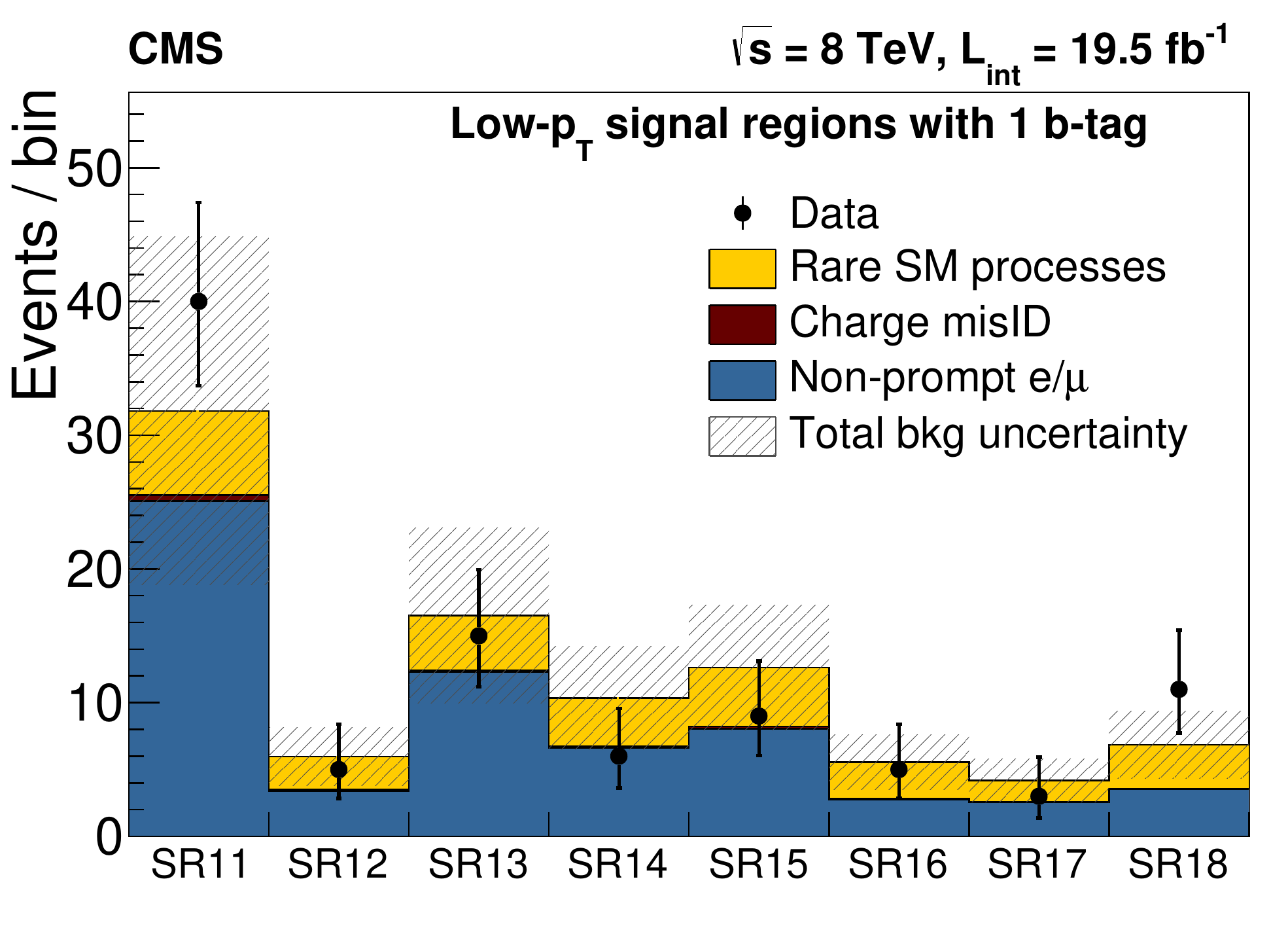} 
        \includegraphics[width=0.49\textwidth]{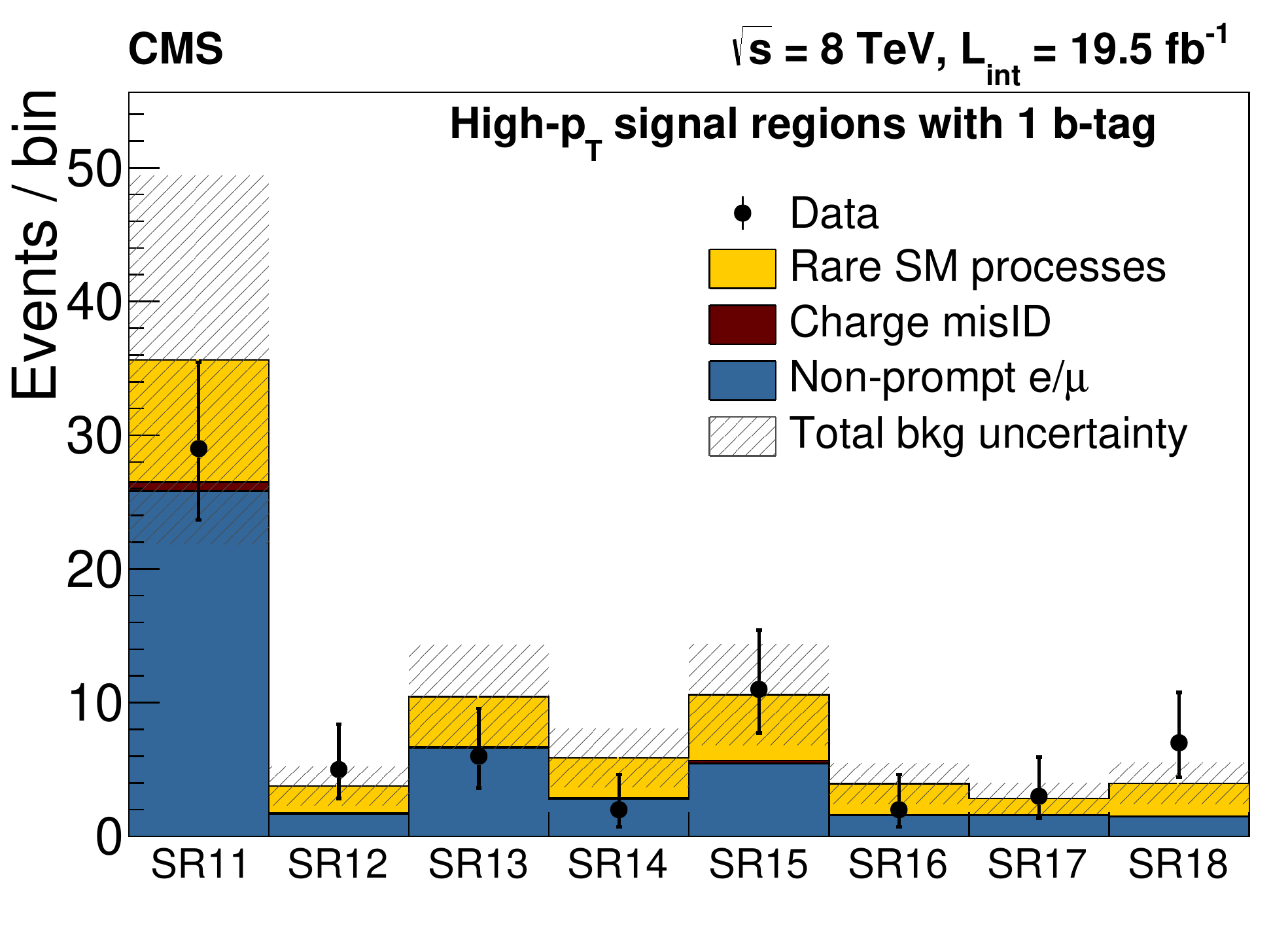}

	\includegraphics[width=0.49\textwidth]{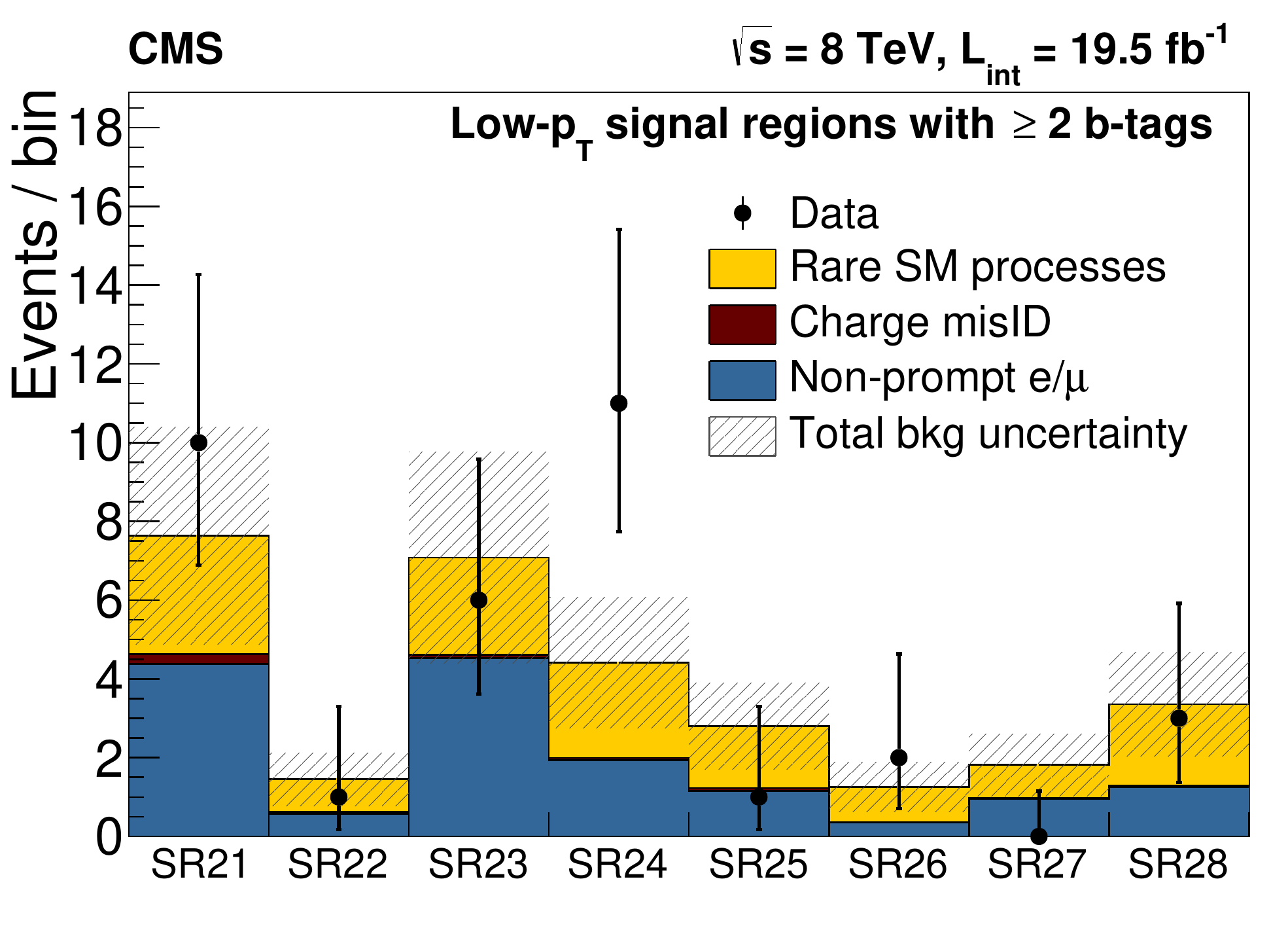} 
        \includegraphics[width=0.49\textwidth]{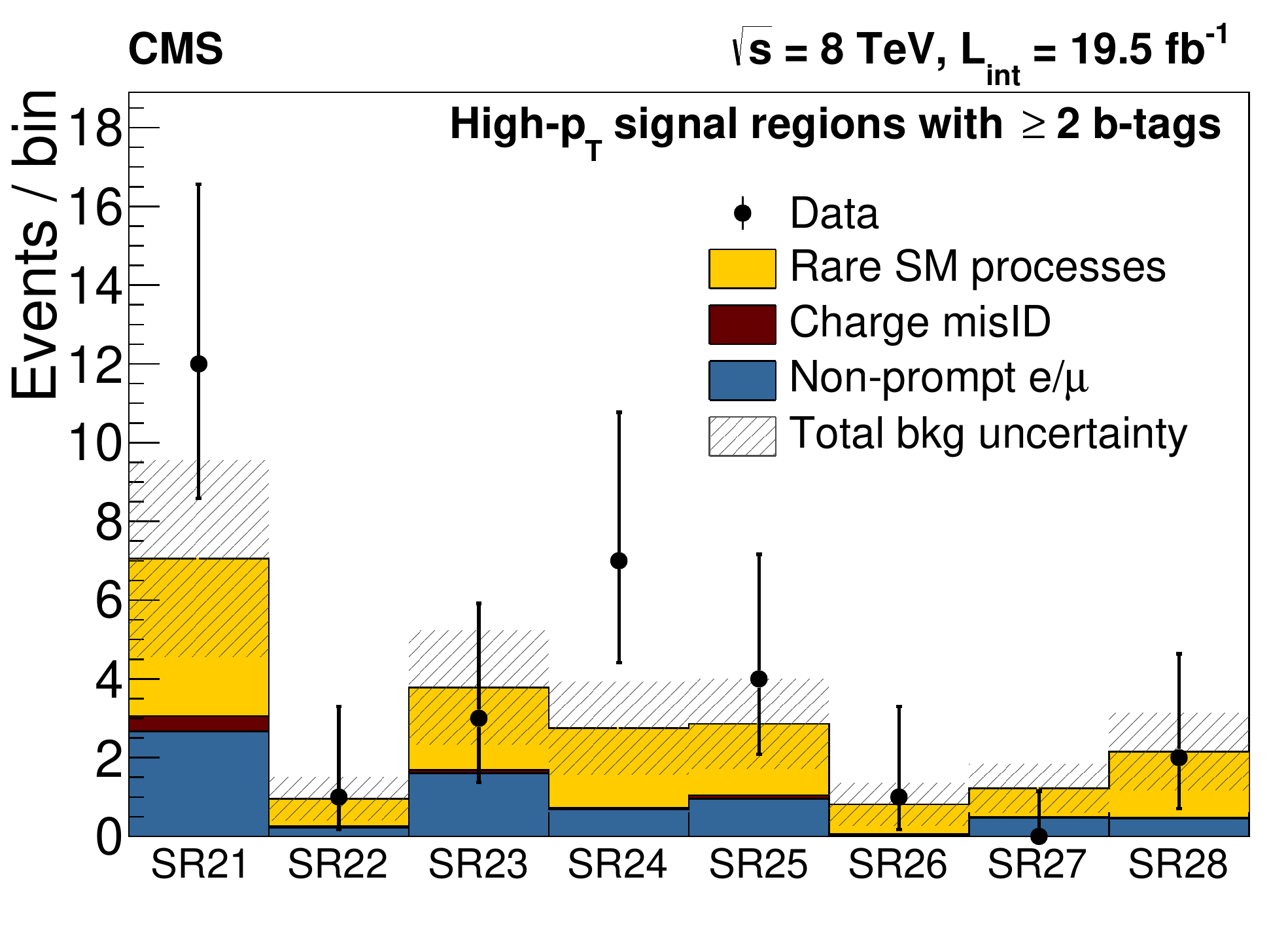}
}

\ifPAS{
	\includegraphics[width=0.49\textwidth]{figs/results/PAS/lowptsummary_regions0.pdf} 
        \includegraphics[width=0.49\textwidth]{figs/results/PAS/highptsummary_regions0.pdf}

	\includegraphics[width=0.49\textwidth]{figs/results/PAS/lowptsummary_regions1.pdf} 
        \includegraphics[width=0.49\textwidth]{figs/results/PAS/highptsummary_regions1.pdf}

	\includegraphics[width=0.49\textwidth]{figs/results/PAS/lowptsummary_regions2.pdf} 
        \includegraphics[width=0.49\textwidth]{figs/results/PAS/highptsummary_regions2.pdf}
}
	\caption{Summary plots showing the predicted background from each source 
	and observed event yields as a function of the SRs in the low-\pt (high-\pt) analysis on  
	left (right).}
	\label{fig:resultsStandardRegions}
	\end{center}
\end{figure}

The observations in each of the final signal regions are presented in Tables~\ref{tab:eventYields} 
and~\ref{tab:eventYieldsSpecial} and in Fig.~\ref{fig:resultsStandardRegions} along with the corresponding 
SM background prediction. 
The contributions of rare SM processes and non-prompt leptons vary among the signal 
regions between 40\% and 60\%, while the charge misidentification background is almost negligible for all signal regions.
The observations are consistent with the background 
expectations within their uncertainties. The p-values~\cite{PDG2012} for each signal region in the low- and high-\pt 
analyses are studied, and are found to be consistent with a uniform distribution between 0 and 1.

\begin{table}[h!]
\begin{center}
\topcaption{Predicted and observed event yields for the low-\pt and high-\pt signal regions.}
\label{tab:eventYields}
\begin{tabular}{c|rclc|rclc}
\hline \hline
\multirow{2}{*}{Region}   & \multicolumn{4}{c|}{Low-\pt}  & \multicolumn{4}{c}{High-\pt}  \\ \cline{2-9}
 & \multicolumn{3}{c}{Expected}  & Observed & \multicolumn{3}{c}{Expected}  & Observed\\ \hline
SR01  & 44 & $\pm$ &16   & 50 & 51 & $\pm$ &18   & 48 \\
SR02  & 12 & $\pm$ &4    & 17 & 9.0 & $\pm$ &3.5 & 11 \\
SR03  & 12 & $\pm$ &5    & 13 & 8.0 & $\pm$ &3.1 & 5  \\
SR04  & 9.1 & $\pm$ &3.4 & 4  & 5.6 & $\pm$ &2.1 & 2  \\
SR05  & 21 & $\pm$ &8    & 22 & 20 & $\pm$ &7    & 12 \\
SR06  & 13 & $\pm$ &5    & 18 & 9 & $\pm$ &4     & 11 \\
SR07  & 3.5 & $\pm$ &1.4 & 2  & 2.4 & $\pm$ &1.0 & 1  \\
SR08  & 5.8 & $\pm$ &2.1 & 4  & 3.6 & $\pm$ &1.5 & 3  \\
&&&&&&&&\\[-0.3cm] \hline &&&&&&&&\\[-0.3cm]
SR11 & 32 & $\pm$ &13   & 40 & 36 & $\pm$ &14   & 29 \\
SR12 & 6.0 & $\pm$ &2.2 & 5  & 3.8 & $\pm$ &1.4 & 5  \\
SR13 & 17 & $\pm$ &7    & 15 & 10 & $\pm$ &4    & 6  \\
SR14 & 10 & $\pm$ &4    & 6  & 5.9 & $\pm$ &2.2 & 2  \\
SR15 & 13 & $\pm$ &5    & 9  & 11 & $\pm$ &4    & 11 \\
SR16 & 5.5 & $\pm$ &2.0 & 5  & 3.9 & $\pm$ &1.5 & 2  \\
SR17 & 4.2 & $\pm$ &1.6 & 3  & 2.8 & $\pm$ &1.1 & 3  \\
SR18 & 6.8 & $\pm$ &2.5 & 11 & 4.0 & $\pm$ &1.5 & 7  \\
&&&&&&&&\\[-0.3cm] \hline &&&&&&&&\\[-0.3cm]
SR21 & 7.6 & $\pm$ &2.8 & 10 & 7.1 & $\pm$ &2.5 & 12 \\
SR22 & 1.5 & $\pm$ &0.7 & 1  & 1.0 & $\pm$ &0.5 & 1  \\
SR23 & 7.1 & $\pm$ &2.7 & 6  & 3.8 & $\pm$ &1.4 & 3  \\
SR24 & 4.4 & $\pm$ &1.7 & 11 & 2.8 & $\pm$ &1.2 & 7  \\
SR25 & 2.8 & $\pm$ &1.1 & 1  & 2.9 & $\pm$ &1.1 & 4  \\
SR26 & 1.3 & $\pm$ &0.6 & 2  & 0.8 & $\pm$ &0.5 & 1  \\
SR27 & 1.8 & $\pm$ &0.8 & 0  & 1.2 & $\pm$ &0.6 & 0  \\
SR28 & 3.4 & $\pm$ &1.3 & 3  & 2.2 & $\pm$ &1.0 & 2  \\
\hline \hline
\end{tabular}
\end{center}
\end{table}

\begin{table}[h!]
\begin{center}
\topcaption{Predicted and observed event yields in the signal regions designed for
  same-sign top-quark pair production and RPV SUSY models.}
\label{tab:eventYieldsSpecial}
\begin{tabular}{l|rcl|c}
\hline \hline
\multicolumn{1}{c|}{SR} & \multicolumn{3}{c|}{Expected}  & Observed \\ \hline
RPV0     & 38  & $\pm$ & 14  & 35  \\
RPV2     & 5.3 & $\pm$ & 2.1 & 5   \\
SStop1   & 160 & $\pm$ & 59  & 152 \\
SStop1++ & 90  & $\pm$ & 32  & 92  \\
SStop2   & 40  & $\pm$ & 13  & 52  \\
SStop2++ & 22  & $\pm$ & 8   & 25  \\
\hline \hline
\end{tabular}
\end{center}
\end{table}

\clearpage

\section{Limits on models of new physics and on rare SM processes}
\label{sec:models}

Given the lack of a significant excess over the expected SM background, the results
of the search are used to derive limits on the parameters of various models of new physics and
to derive limits on the cross sections of rare SM processes.
The 95\% confidence level (CL) upper limits  on the signal yields are calculated
using the LHC-type CL$_\mathrm{s}$ method~\cite{Read:2002hq,Junk:1999kv,ATL-PHYS-PUB-2011-011}.
Lognormal nuisance parameters are used for the signal (Table~\ref{tab:signalsys}) and background estimate (Tables~\ref{tab:eventYields} and~\ref{tab:eventYieldsSpecial}) uncertainties.
For each model considered, limits are obtained by performing a statistical
combination of the most sensitive signal regions.

The signal regions used to set limits on the new physics models explored in this paper are given in Table~\ref{tab:SRused}.

The number of events that are expected to satisfy the selection for a given signal model
is obtained from MC simulation. The uncertainties for the event yields are computed as described in Section~\ref{sec:eff}.
For a given signal region, the different sources of uncertainties in the signal acceptance
are considered to be uncorrelated, with correlations across signal regions taken into account.
The uncertainties in the total background across the signal regions are considered to be fully correlated.

\begin{figure}[t]
\begin{center}
\includegraphics[width=0.45\linewidth]{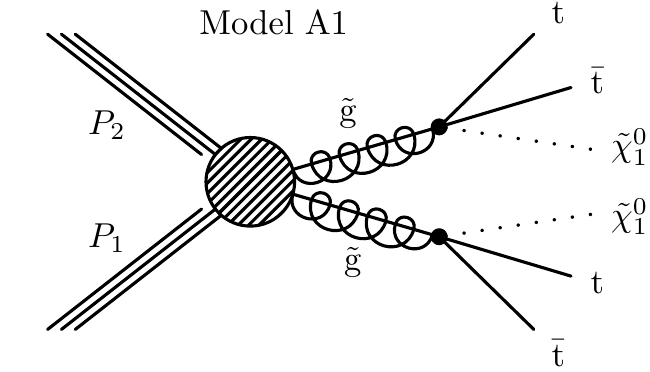} \hfill
\includegraphics[width=0.45\linewidth]{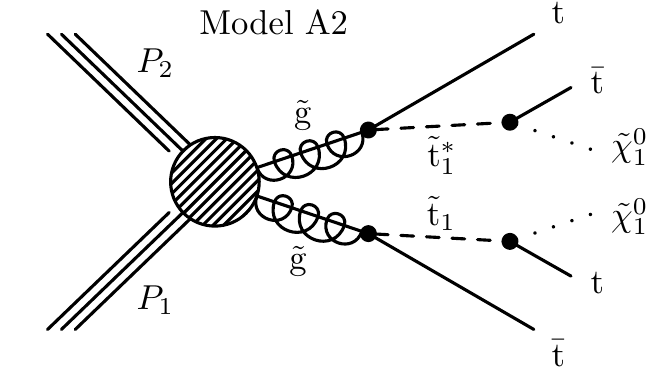}
\includegraphics[width=0.45\linewidth]{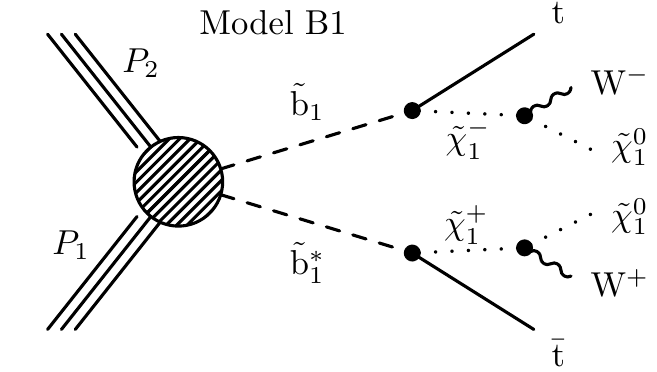} \hfill
\includegraphics[width=0.45\linewidth]{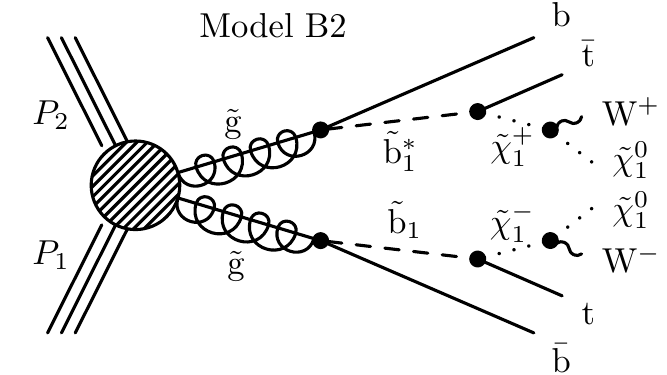}
\includegraphics[width=0.45\linewidth]{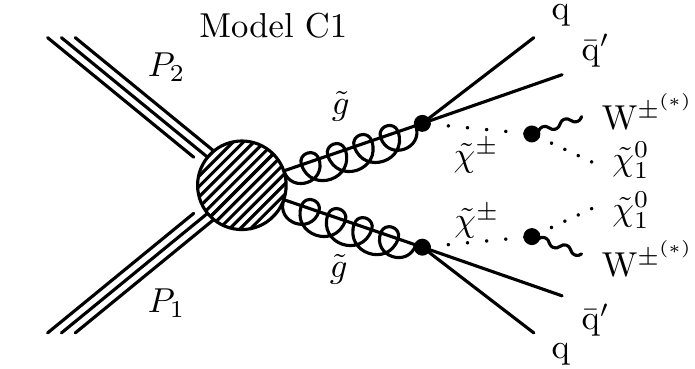} \hfill
\includegraphics[width=0.45\linewidth]{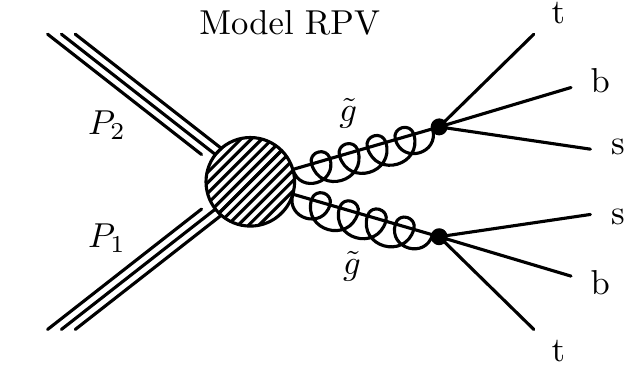}
\caption{Diagrams for the six SUSY models considered (A1, A2, B1, B2, C1, and RPV).}
\label{fig:FD}
\end{center}
\end{figure}

\begin{table}[h]
\begin{center}
\topcaption{\label{tab:SRused} Signal regions used for limit setting for the new physics models considered in this analysis.}
\begin{tabular}{c|c|c|c}
\hline\hline
Model & Constraints on parameters                & Analysis   & Signal regions used \\ \hline\hline
A1 &                                   & high-$\pt$ & 21--28 \\
A2 & m$_{\chiz_1}$ = 50\GeV              & high-$\pt$ & 21--28 \\
B1 & m$_{\chiz_1}$ = 50\GeV              & high-$\pt$ & 11--18, 21--28 \\
B1 & $\text{m}_{\chiz_1}/\text{m}_{\chipm_1} = 0.5$ & high-$\pt$ & 11--18, 21--28 \\
B1 & $\text{m}_{\chiz_1}/\text{m}_{\chipm_1} = 0.8$ & low-$\pt$ & 11--18, 21--28 \\
B2 & m$_{\chiz_1}$ = 50\GeV, m$_{\chipm_1}$ = 150\GeV & high-$\pt$ & 21--28 \\
B2 & m$_{\chiz_1}$ = 50\GeV, m$_{\chipm_1}$ = 300\GeV & high-$\pt$ & 21--28 \\
C1 & $m_{\chipm_1}$ = $0.5m_{\chiz_1}$ + $0.5m_{\sGlu}$ & high-$\pt$ & 01--08 \\
C1 & $m_{\chipm_1}$ = $0.8m_{\chiz_1}$ + $0.2m_{\sGlu}$ & low-$\pt$ & 01--08 \\
RPV &  & high-$\pt$ & RPV2 \\
$\Pp \Pp \to \cPqt \cPqt,~\cPaqt \cPaqt$ &  & high-$\pt$ & SStop1, SStop2 \\
$\Pp \Pp \to \cPqt \cPqt$                 &  & high-$\pt$ & SStop1++, SStop2++\\
$\Pp \Pp \to \cPqt \cPqt \cPaqt \cPaqt$ &  & high-$\pt$ & 21--28 \\
\hline\hline
\end{tabular}
\end{center}
\end{table}

First, we present limits on the parameter spaces of various
$R$-parity-conserving simplified SUSY models~\cite{T1tttt}. The exclusion contours are obtained
with the gluino or bottom-squark pair production cross sections at the NLO+NLL (i.e. next-to-leading-logarithm) accuracy that are calculated
in the limit where other sparticles are heavy enough to be
decoupled~\cite{bib-nlo-nll-01,bib-nlo-nll-02,bib-nlo-nll-03,bib-nlo-nll-04,bib-nlo-nll-05,Kramer:2012bx}.
The production of SUSY particles and the decay chains under consideration are shown schematically in
Fig.~\ref{fig:FD}.

\begin{figure}[tp]
\begin{center}
\ifPAPER{
\includegraphics[width=0.46\linewidth]{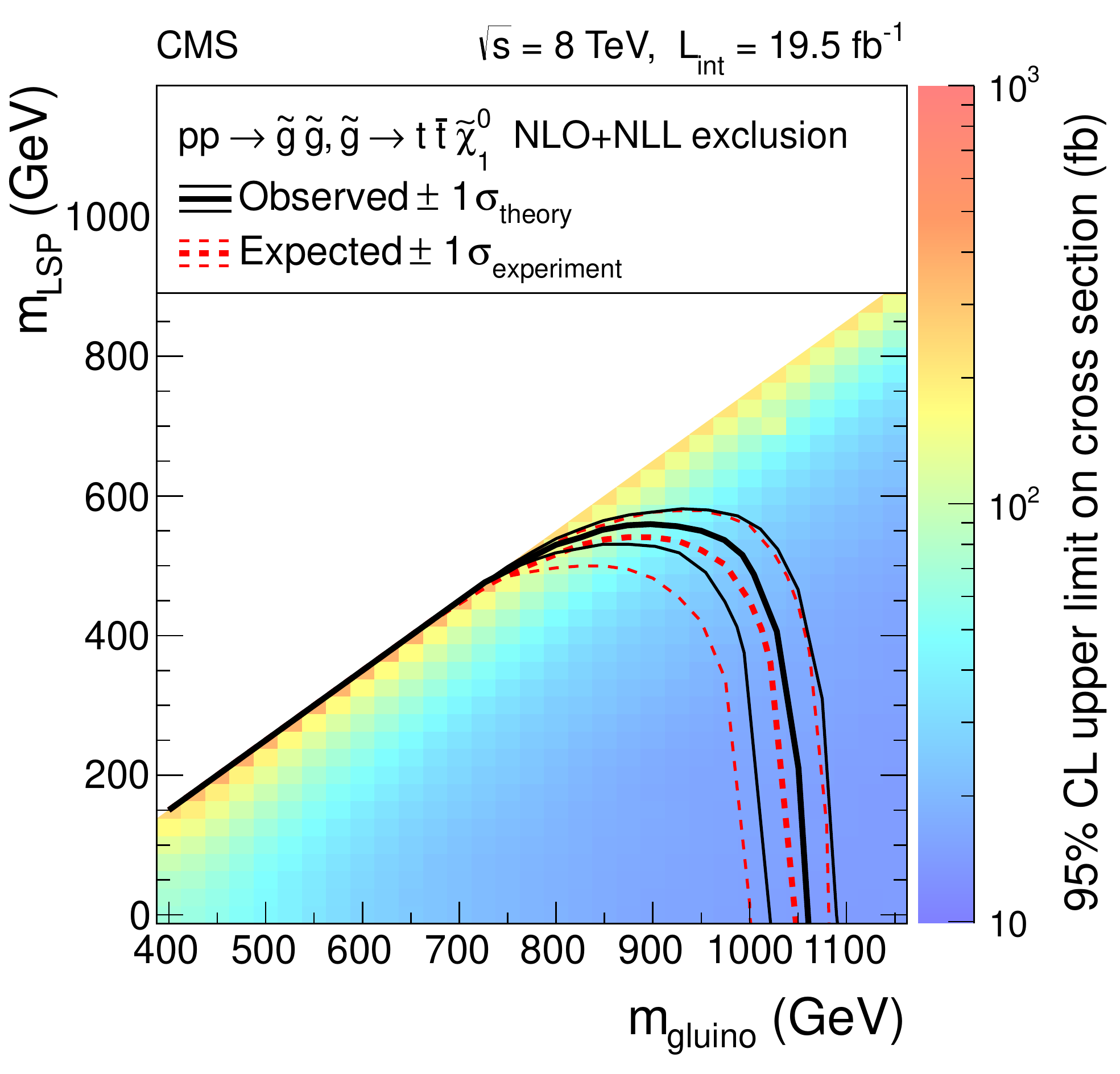}
\includegraphics[width=0.46\linewidth]{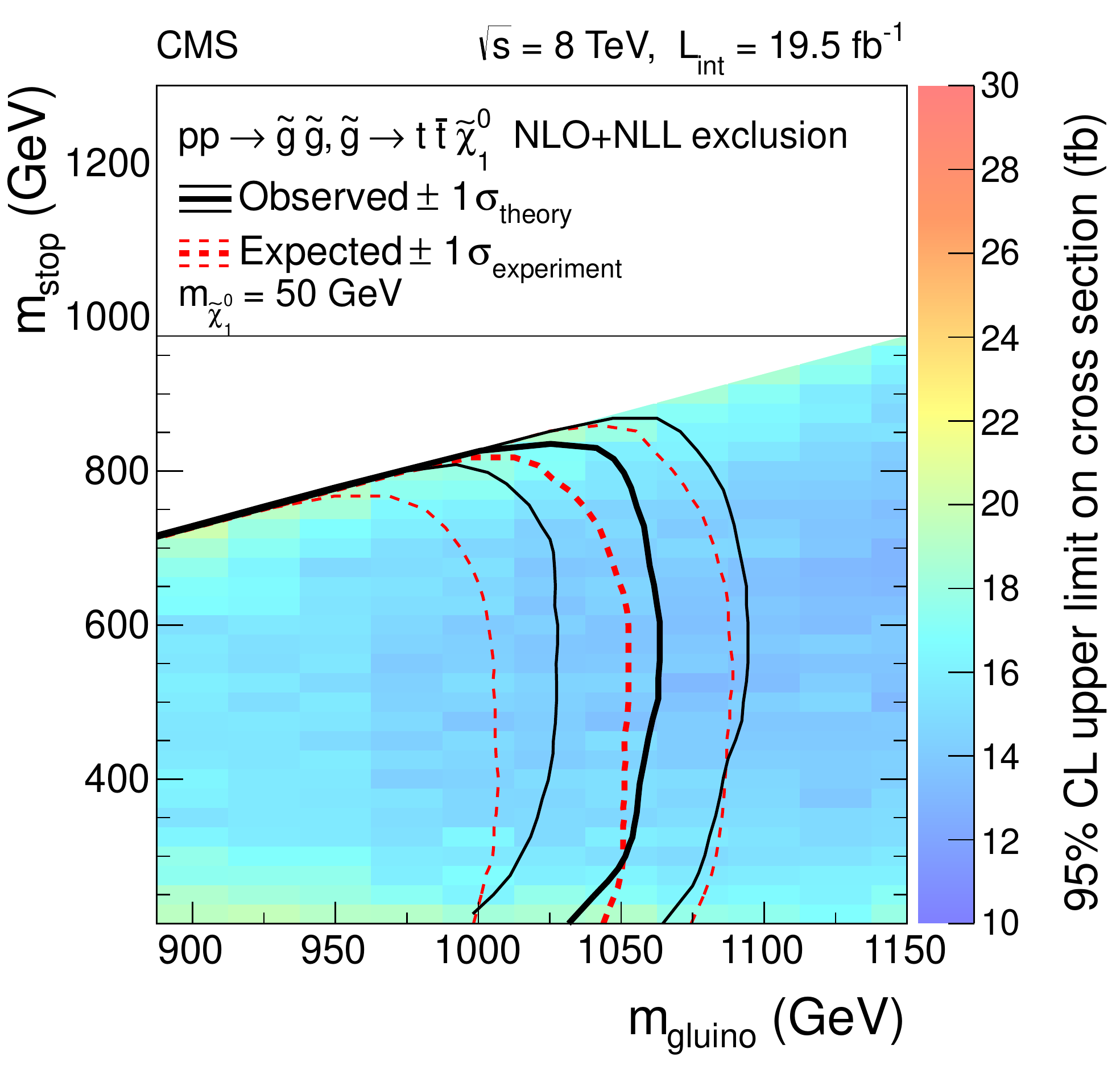}
}
\ifPAS{
\includegraphics[width=0.46\linewidth]{figs/limits/PAS/T1.pdf}
\includegraphics[width=0.46\linewidth]{figs/limits/PAS/T5.pdf}
}
\caption{Exclusion regions at 95\% CL in the planes of
(left) $m(\chiz_1)$ versus $m(\sGlu)$ (model A1), and (right)
$m(\sTop_1)$ versus $m(\sGlu)$ (model A2).
The excluded regions are those within the kinematic boundaries and to
the left of the curves. The effects of the theoretical uncertainties in
the NLO+NLL calculations of the production
cross sections~\cite{Kramer:2012bx} are indicated by the thin black curves; the expected limits and their $\pm$1
standard-deviation variations are shown by the dashed red curves.
\label{fig:T1tttt}}
\end{center}
\end{figure}

Scenarios A1 and A2 represent models of gluino pair production
resulting in the $\cPqt \cPqt \cPaqt \cPaqt \chiz_1 \chiz_1$ final
state, where $\chiz_1$ is the lightest neutralino~\cite{stopVirtual,stopVirtualPRD,T1tttt,wacker,naturalness4}.
In model A1, the gluino undergoes a three-body decay
$\sGlu \to \ttbar \chiz_1$ mediated by an off-shell top squark.
In model A2, the gluino decays to a top quark and a top anti-squark,
with the on-shell anti-squark further decaying into a top anti-quark and a neutralino.
Both of these models produce four on-shell W bosons and four b quarks.
Therefore, search regions SR21--SR28, which require at least two b-tagged jets
and high-\pt leptons, are used to derive the limits on the parameters of these models;
the region with the best sensitivity is SR28.
The 95\% CL upper limits on the cross section times branching fraction, as well as
the exclusion contours, are shown in Fig.~\ref{fig:T1tttt}. For model A1, the results are presented
as a function of gluino mass and $\chiz_1$ mass, and for model A2
as a function of gluino mass and top squark mass with the $\chiz_1$ mass set to 50\GeV.
In model A2, the limits do not depend on the top squark or $\chiz_1$ masses
provided that there is sufficient phase space to produce on-shell top quarks
with a moderate boost in the decay of both the gluino and the top squark.
This range extends to approximately $600$\GeV for the $\chiz_1$ mass.

\begin{figure}[tp]
\begin{center}
\ifPAPER{
\includegraphics[width=0.46\linewidth]{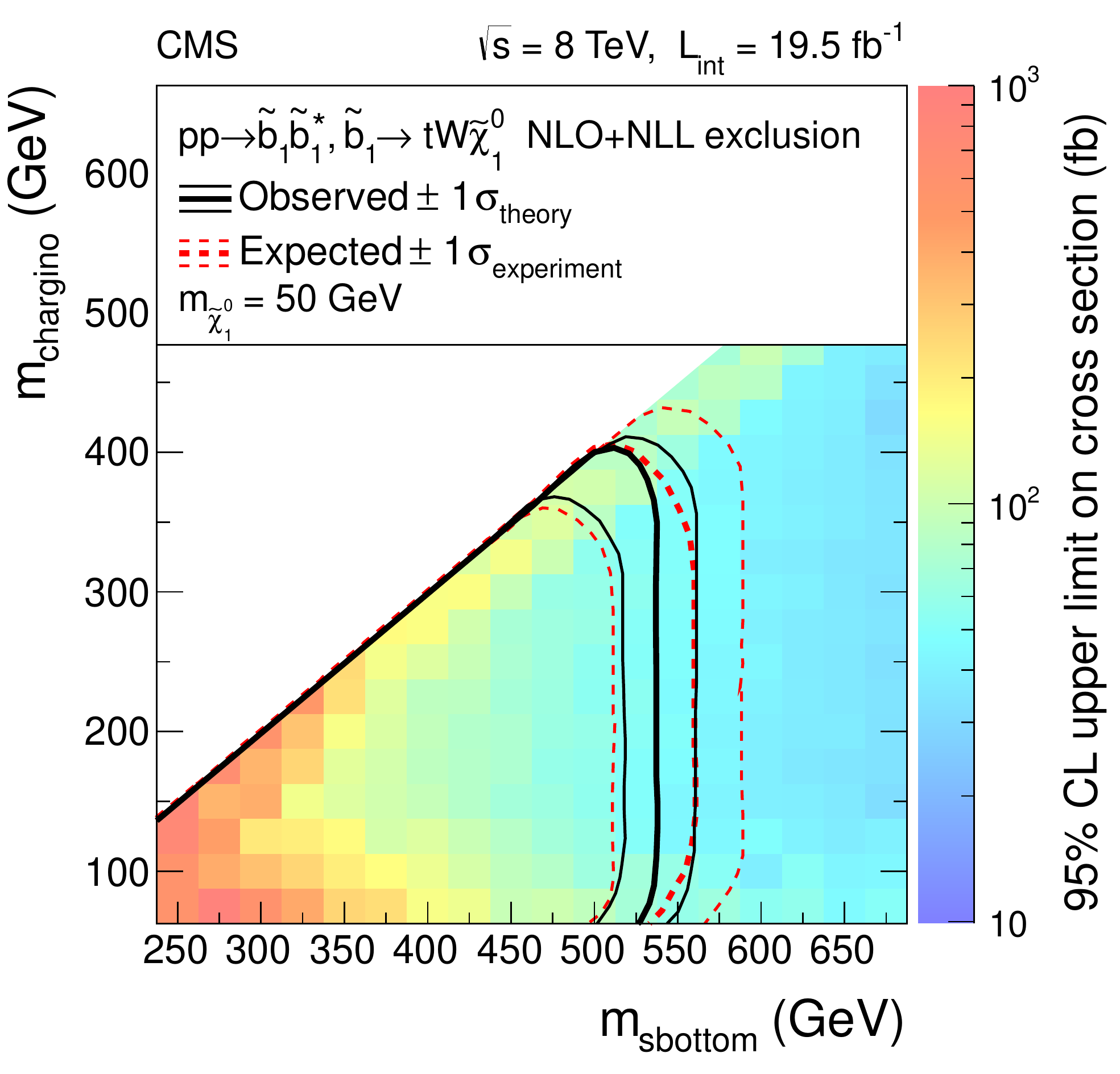} \\
\includegraphics[width=0.46\linewidth]{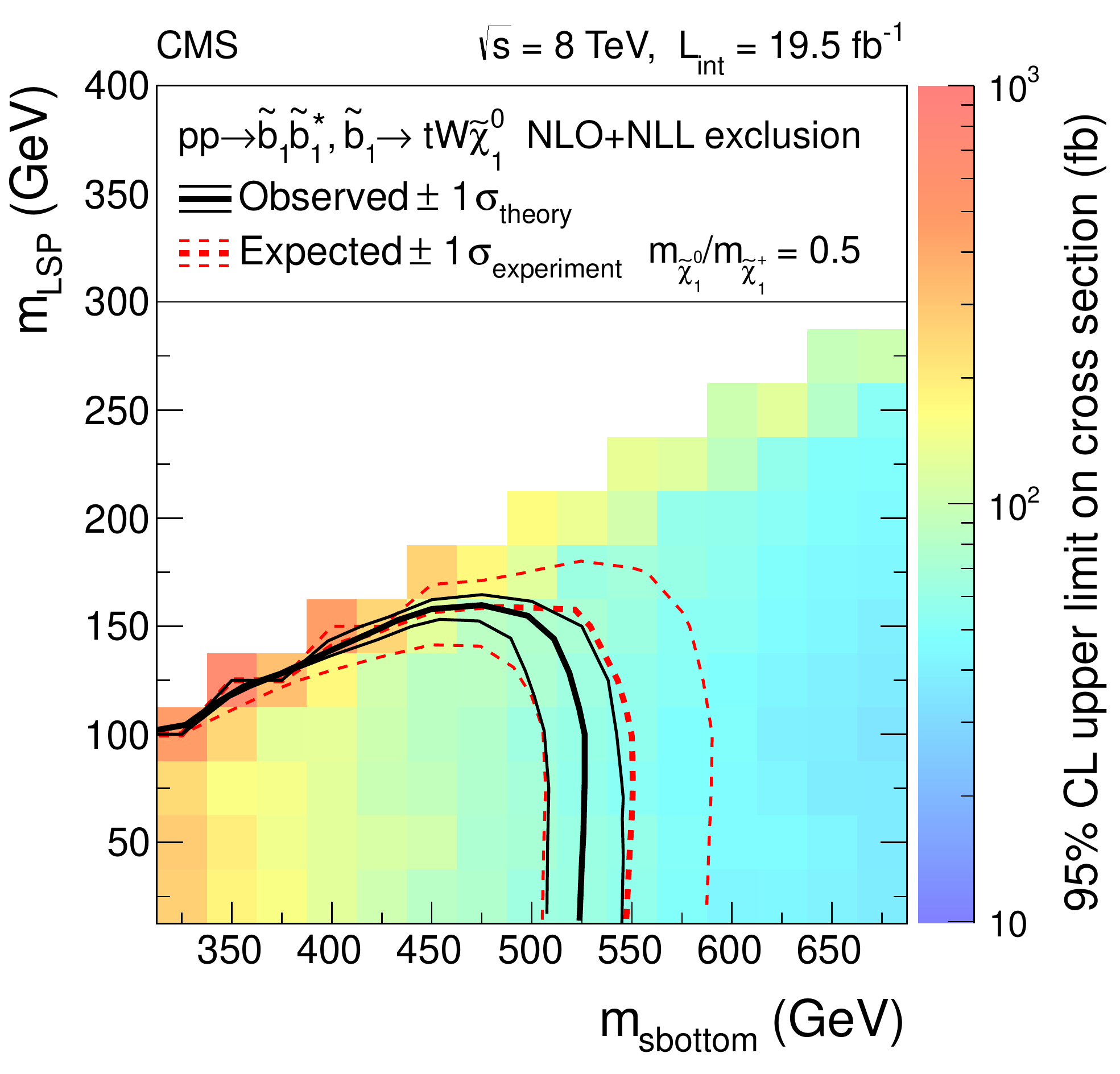}
\includegraphics[width=0.46\linewidth]{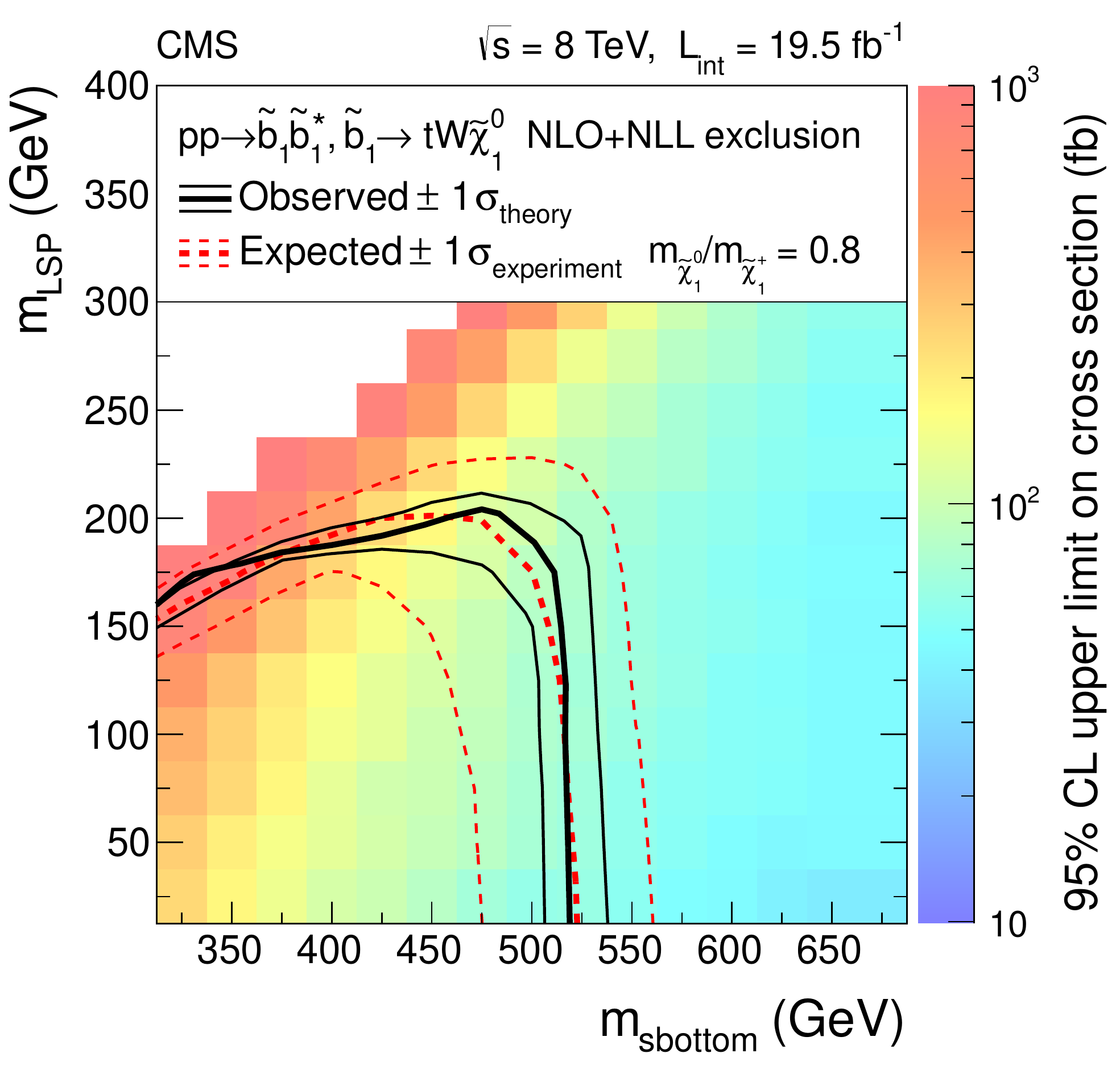}
\includegraphics[width=0.46\linewidth]{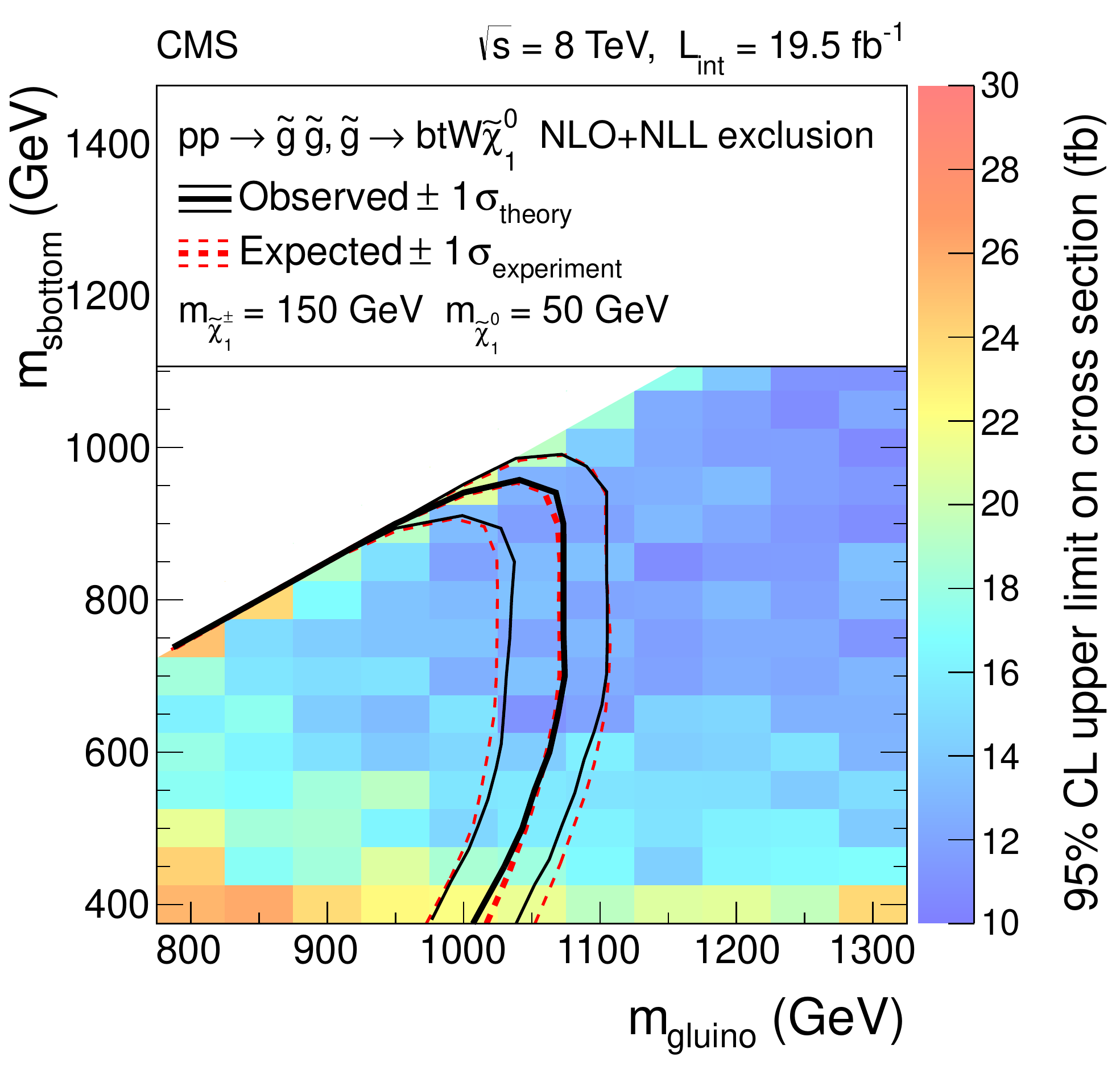}
\includegraphics[width=0.46\linewidth]{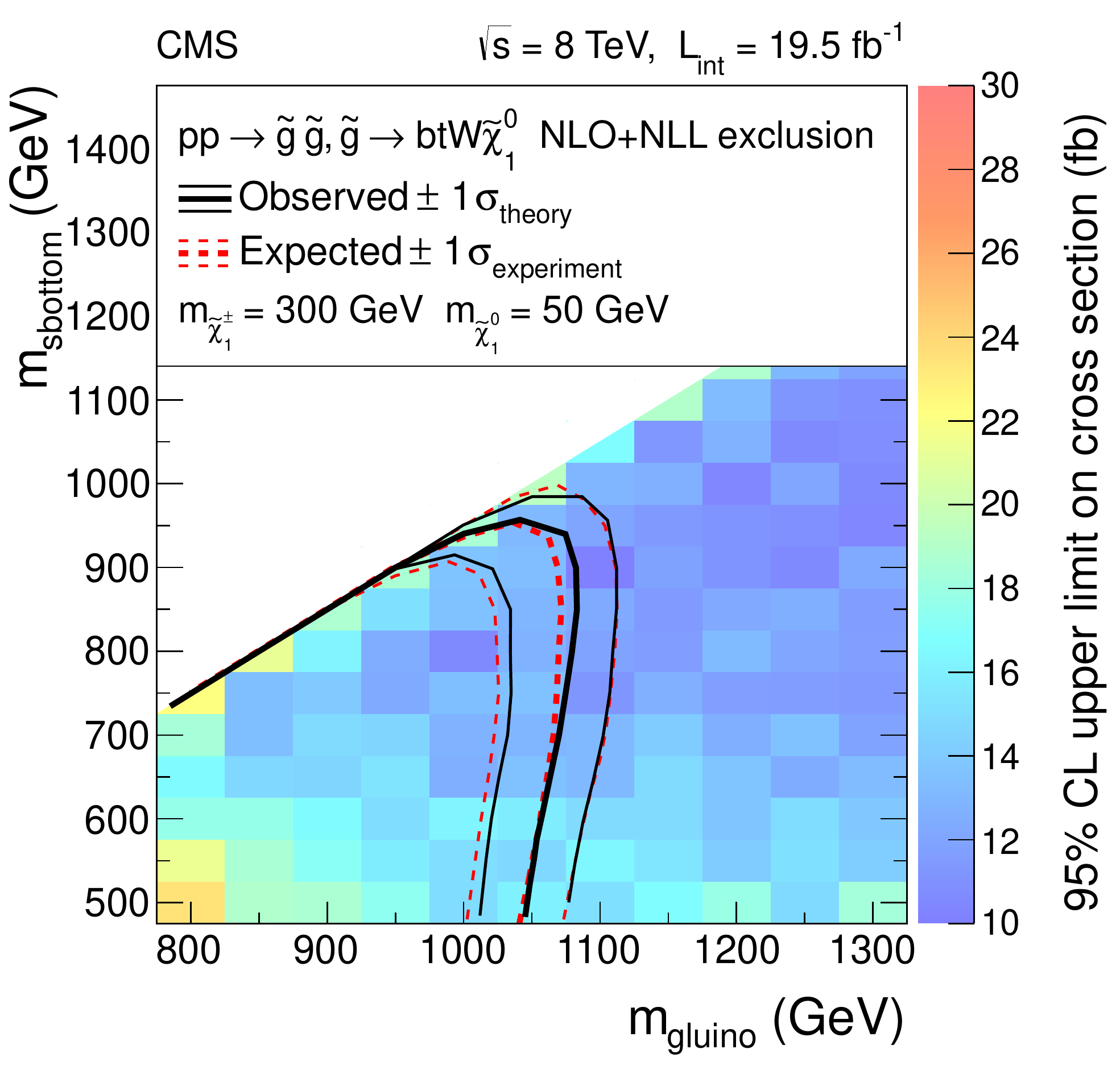}
}
\ifPAS{
\includegraphics[width=0.46\linewidth]{figs/limits/PAS/T6.pdf} \\
\includegraphics[width=0.46\linewidth]{figs/limits/PAS/T6x05.pdf}
\includegraphics[width=0.46\linewidth]{figs/limits/PAS/T6x08.pdf}
\includegraphics[width=0.46\linewidth]{figs/limits/PAS/T7150.pdf}
\includegraphics[width=0.46\linewidth]{figs/limits/PAS/T7300.pdf}
}
\caption{Exclusion regions at 95\% CL in the planes of
(top and center) $m(\chipm_1)$ versus $m(\sBot_1)$ and $m(\chiz_1)$ versus $m(\sBot_1)$ (model B1), and
(bottom) $m(\sBot_1)$ versus $m(\sGlu)$ (model B2). The convention for the exclusion curves is the same as in Fig.~\ref{fig:T1tttt}.
\label{fig:T6bbWW}}
\end{center}
\end{figure}

Model B1 is a model of bottom-squark pair production, followed
by one of the most likely decay modes of the bottom squark,
$\sBot_1 \to \cPqt \chim_1$ with $\chim_1 \to \PWm \chiz_1$, where $\sBot_1$
and $\chim_1$ represent the lightest bottom squark and lightest chargino, respectively.
We consider three cases in this decay mode. We either set the $\chiz_1$
mass to 50\GeV and present the limits in the ($m_{\chipm_1}$, $m_{\sBot_1}$) plane,
or consider the ($m_{\chiz_1}$, $m_{\sBot_1}$) plane with the mass of the chargino set according to
${m}_{\chiz}/{m}_{\chipm_1} = 0.5$ or ${m}_{\chiz}/{m}_{\chipm_1} = 0.8$.
The values 0.5 and 0.8 are representative choices that determine
whether the top quark and W boson are on-shell or off-shell, which has a direct impact on
the sensitivity of the analysis in this model.
The limits for this model, obtained using search regions SR11 to SR28,
are presented in Fig.~\ref{fig:T6bbWW}. For ${m}_{\chiz}/{m}_{\chipm_1} = 0.8$, the
low-\pt lepton selection is used, while high-\pt leptons are used for the other two scenarios.
SR28 is again the most sensitive signal region, followed by the regions requiring
one b-tagged jet: SR18, SR15, and SR13.

Model B2 consists of gluino pair production followed by $\sGlu \to \sBot_1 \cPaqb$.
The gluino decay modes in models A1 and A2 are expected to be dominant
if the top squark is the lightest squark.
Conversely, if the bottom squark is the lightest, the decay mode
in model B2 would be the most probable. The limits on this model, calculated using search regions
SR21--SR28 and the high-\pt lepton selection, are presented in Fig.~\ref{fig:T6bbWW} as a function
of $m(\sBot_1)$ and  $m(\sGlu)$ for two fixed masses of $m_{\chipm_1}$, 150 and 300\GeV.
The region with the largest sensitivity to this model is SR28.

\begin{figure}[tp]
\begin{center}
\ifPAPER{
\includegraphics[width=0.46\linewidth]{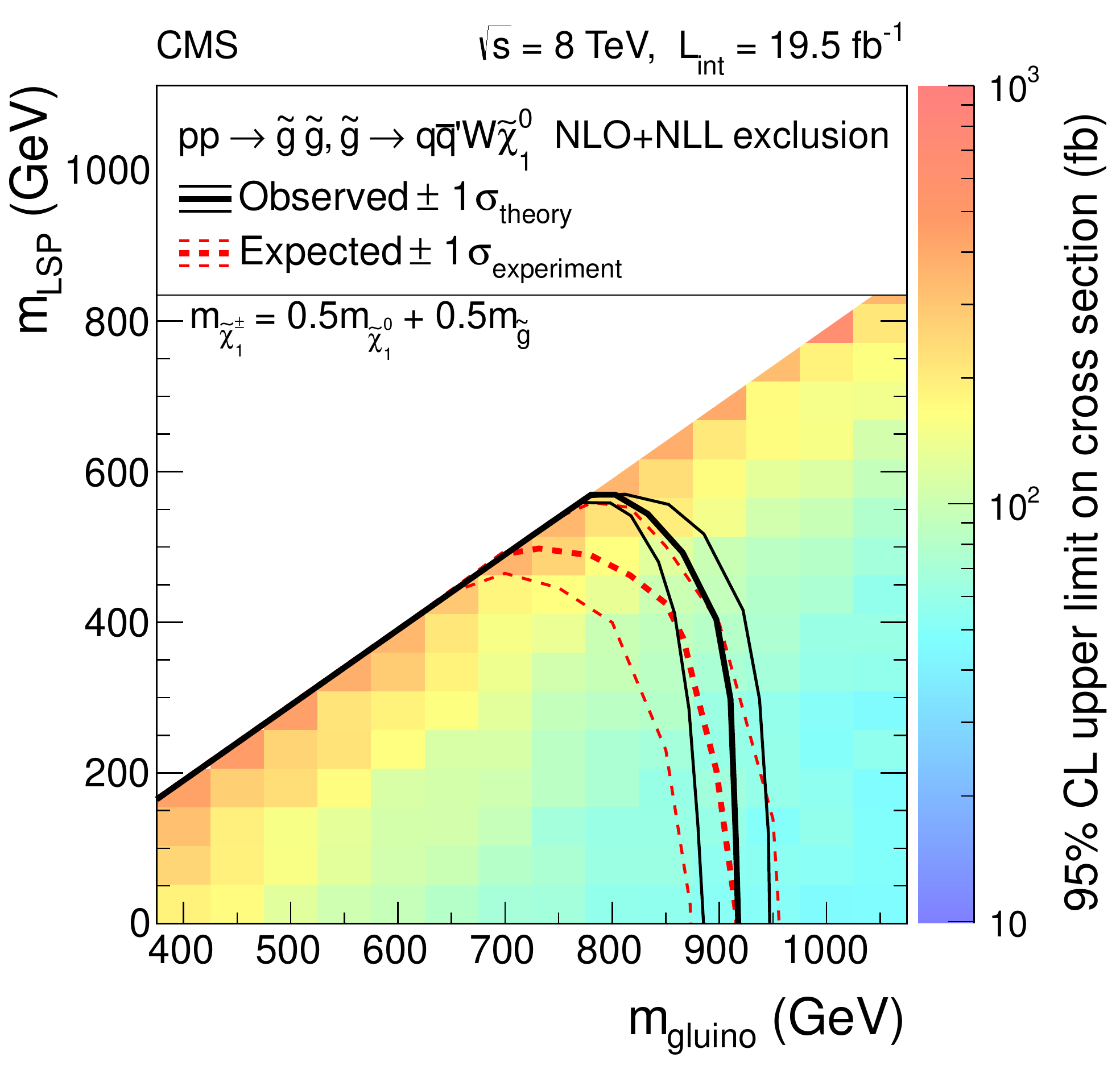}
\includegraphics[width=0.46\linewidth]{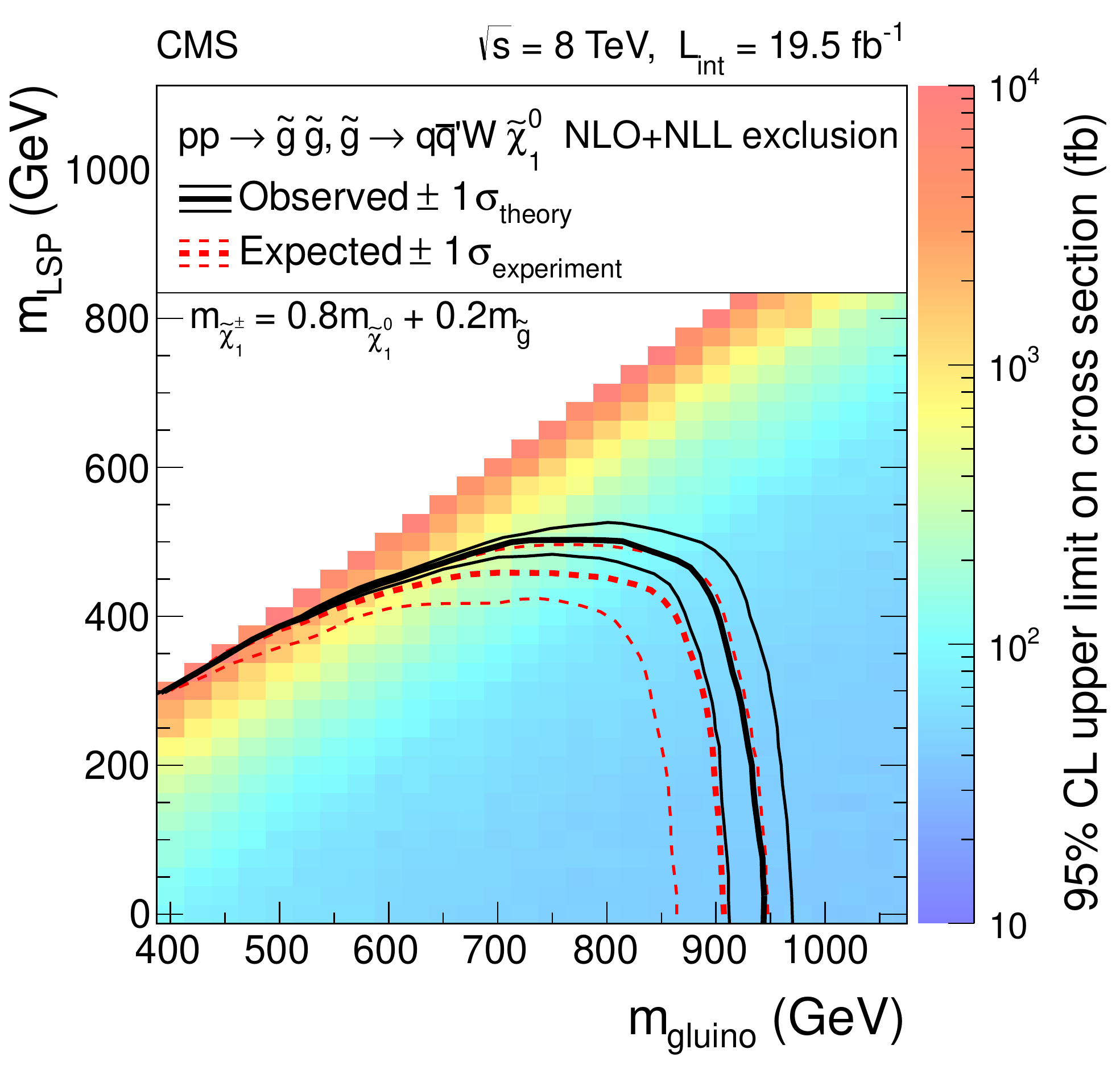}
}
\ifPAS{
\includegraphics[width=0.46\linewidth]{figs/limits/PAS/TVV.pdf}
\includegraphics[width=0.46\linewidth]{figs/limits/PAS/Tlnu.pdf}
}
\caption{Exclusion regions at 95\% CL in the planes of
$m(\chiz_1)$ versus $m(\sGlu)$ for two different values of chargino mass (model C1). The convention for the exclusion curves is the same as in Fig.~\ref{fig:T1tttt}.
\label{fig:T5Lnu}}
\end{center}
\end{figure}

Model C1 is based on the production of a gluino pair
where each gluino decays to light quarks and a chargino via heavy virtual squarks:
$\sGlu\to\Pq\Paq'\chipm_1$, $\chipm_1 \to \text{W}^{(*)}\chiz_1$.
The decay is charge-symmetric, resulting in an equal fraction of same-sign and opposite-sign W boson pairs
in the final state.
In this model there are three parameters: $m_{\sGlu}$, $m_{\chipm_1}$, and $m_{\chiz_1}$.
Signal samples are produced for each bin in the
$(m_{\chiz_1},m_{\sGlu})$ plane. Chargino mass is defined through a parameter $x$ as
$m_{\chipm_1}$ = $xm_{\chiz_1}$ + $(1-x)m_{\sGlu}$. In the limit $x\rightarrow0$,
there is no observable hadronic activity in the event. At the other extreme, $x\rightarrow1$,
the chargino and LSP are degenerate and the chargino decays through
an off-shell W boson yielding very soft leptons. In either cases, the analysis loses sensitivity.
For intermediate values of the parameter $x$, the W boson is either on- or off-shell depending on
the values of $m_{\chiz_1}$ and $m_{\sGlu}$, giving rise to either
high- or low-\pt leptons. We examine $x$ values of 0.5 and 0.8. The former value ensures that the W boson
is on-shell in the sparticle mass range considered, while the latter yields mostly off-shell W bosons.
In this model, no enrichment of heavy-flavour jets is expected.
Therefore, the search regions SR01--SR08, with both the low- and high-\pt lepton selection, are
used for cross section upper limit calculation. The limits are presented in Fig.~\ref{fig:T5Lnu}.
In this model, gluino masses up to 900\GeV are probed. Most of the
sensitivity to this model is obtained from signal region SR08.

\begin{figure}[tp]
\begin{center}
\ifPAPER{
\includegraphics[width=0.60\linewidth]{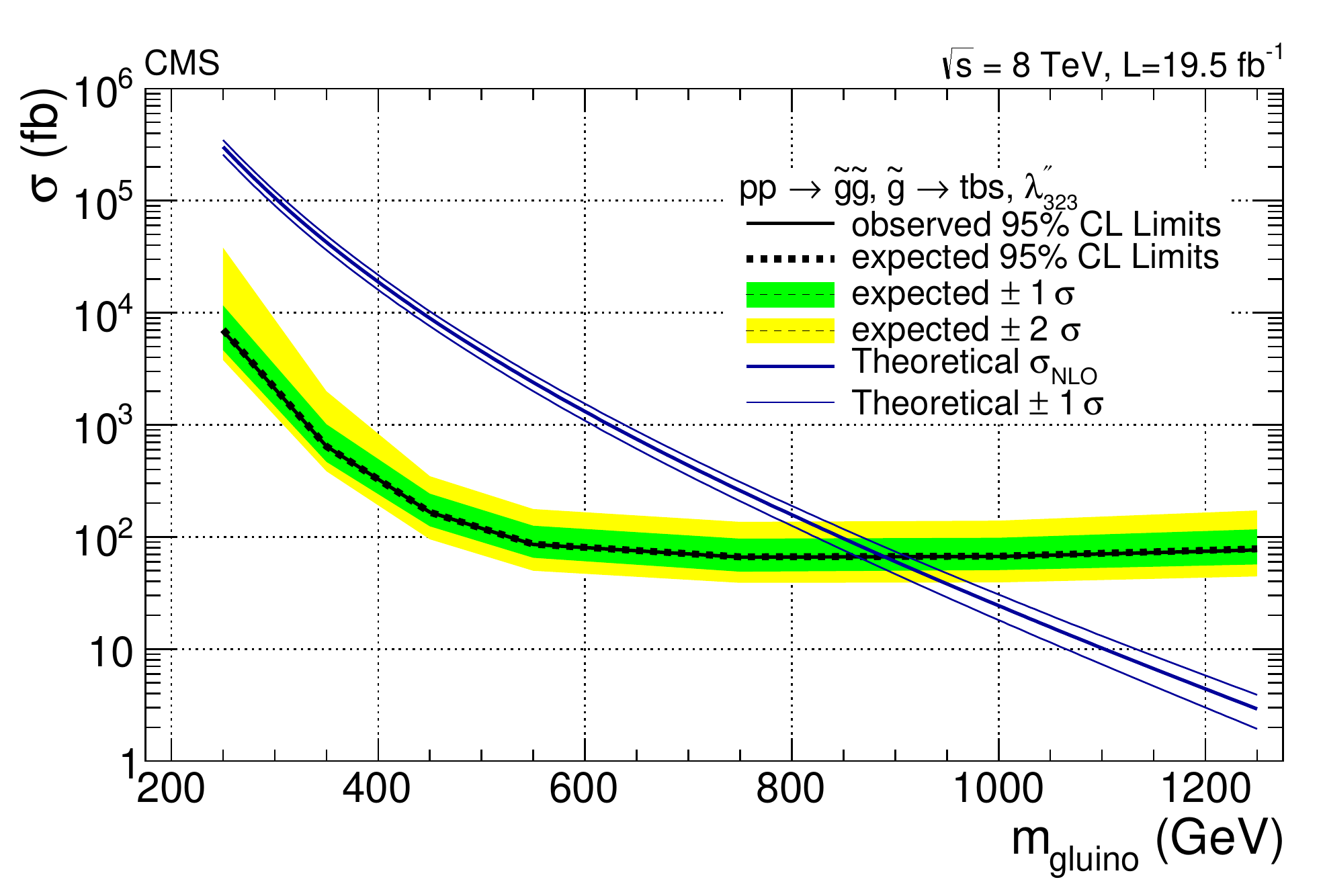}
}
\ifPAS{
\includegraphics[width=0.60\linewidth]{figs/limits/PAS/RPV.pdf}
}
\caption{95\% CL upper limit on the gluino production cross section for an RPV simplified model,
$\Pp \Pp \to \sGlu \sGlu, \sGlu \to \cPqt\cPqb\cPqs (\cPaqt\cPaqb\cPaqs)$.
\label{fig:RPVlimit}}
\end{center}
\end{figure}

These results extend the sensitivity obtained in the previous analysis~\cite{sspaper8TeV10fb}
on gluino and sbottom masses.
For the gluino-initiated models (A1, A2, B2, and C1), we probe gluinos
with masses up to about 1050\GeV, with relatively small dependence on the
details of the models. This is because the limits are driven by the
common gluino pair production cross section.
In the case of the direct bottom-squark pair production, model
B1, our search shows sensitivity for bottom-squark masses up to about $500$\GeV.

These models are also probed by other CMS new physics searches
in different decay modes. Other searches are usually interpreted in the context
of model A1 but not A2, B1, or B2.
For model A1, the limits given here are
complementary to the limits from the
searches presented in Refs.~\cite{maria,Chatrchyan:2012wa,Chatrchyan:2013lya,Chatrchyan:2013wxa}.
In particular, they are
less stringent at low $m(\chiz_1)$ but more stringent at high $m(\chiz_1)$.
A similar conclusion applies to model A2, since the final state is the same. For
bottom-squark pair production, limits on $m(\sBot_1)$ of
about 600\GeV have been presented~\cite{Chatrchyan:2013lya}, but assuming
the decay mode $\sBot_1 \to \cPqb \chiz_1$
instead of the model B1 mode $\sBot_1 \to \cPqt \chim_1$ considered here.
Comparable limits for model A1, as well as for similar models with
top and bottom quarks from gluino decays, have been
reported by the ATLAS Collaboration~\cite{ATLAS:2012ai,ATLAS:2012ah,ATLAS:2012pq,Aad:2012naa}.

A single RPV scenario is considered in this analysis, one in which gluino pair production is followed
by the decay of each gluino to three quarks, as is favoured in the SUSY model
with minimal flavour violation~\cite{Csaki:2011ge}:
$\sGlu\to\cPqt\cPqb\cPqs (\cPaqt\cPaqb\cPaqs)$ (model RPV). Such decays lead to
same-sign W-boson pairs in the final state in 50\% of the cases. Compared with the decays
$\sGlu\to\cPqt\cPqs\cPqd (\cPaqt\cPaqs\cPaqd)$, which also yield same-sign W-boson pairs,
the mode considered profits from having two extra b quarks in the final state,
resulting in a higher signal selection efficiency. The model is governed by one parameter
($m_{\sGlu}$), which dictates the production cross section and the final state
kinematics. The dedicated search region RPV2 with the high-$\pt$ lepton selection
is used to place an upper limit on the production cross section.
The result is
shown in Fig.~\ref{fig:RPVlimit}. In this scenario, the gluino mass is probed
up to approximately 900\GeV.

The results for the signal regions SStop1, SStop1++, SStop2, and SStop2++
are used to set limits on
the cross section for same-sign top-quark pair production,
$\sigma(\Pp \Pp \to \cPqt \cPqt,~\cPaqt \cPaqt)$ from  SStop1 and SStop2, and
$\sigma(\Pp \Pp \to \cPqt \cPqt)$ from SStop1++ and SStop2++.
Here
$\sigma(\Pp \Pp \to \cPqt \cPqt,~\cPaqt \cPaqt)$ is shorthand
for the sum
$\sigma(\Pp \Pp \to \cPqt \cPqt)+\sigma(\Pp \Pp \to \cPaqt \cPaqt)$.
These limits are calculated
using an acceptance obtained from simulated $\Pp \Pp \to \cPqt \cPaqt$ events and an opposite-sign selection.
This acceptance, including branching fractions, is $0.43\%$ ($0.26\%$) for the SStop1 (SStop2) search region.
The relative uncertainty in this acceptance is $14\%$.
The observed upper limits are $\sigma(\Pp \Pp \to \cPqt \cPqt,~\cPaqt \cPaqt) < 720$\unit{fb}
and $\sigma(\Pp \Pp \to \cPqt \cPqt) < 370$\unit{fb} at 95\% CL.
The median expected limits are $470^{+180}_{-110}$\unit{fb} and $310^{+110}_{-80}$\unit{fb}, respectively.

Similarly, the results from signal regions SR21--SR28 with the high-\pt lepton selection are used
to set limits on the SM cross section for quadruple top-quark production. The observed upper limit
is $\sigma(\Pp \Pp \to \cPqt \cPqt \cPaqt \cPaqt) < 49$\unit{fb} at 95\% CL, compared
to a median expected limit of $36^{+16}_{-9}$\unit{fb}. The SM cross section as computed with the MC$\at$NLO
program~\cite{Frixione:2002ik} is $\sigma_{\rm SM} = 0.914\pm 0.005$\unit{fb}.
The most sensitive signal regions, SR24 and SR28, have a signal acceptance
of 0.52\% and 0.49\%, respectively, with relative uncertainties of 13\% and 17\%.

\section{Information for additional model testing}
\label{sec:outreach}

We have described a signature-based search that finds no evidence
for physics beyond the SM.  In Section~\ref{sec:models}, the results
are used to place bounds on the parameters of a number of models of new physics.
Here, additional information is presented that can be used to confront other
models of new physics in an approximate way through MC generator-level studies.
The expected numbers of events can then be compared with an upper limit on the number of signal events
that can be obtained using
inputs from Tables~\ref{tab:eventYields}~and~\ref{tab:eventYieldsSpecial}
and a signal acceptance uncertainty estimated from the generator-level studies.

\begin{figure}[h]
\begin{center}
\includegraphics[width=0.48\linewidth]{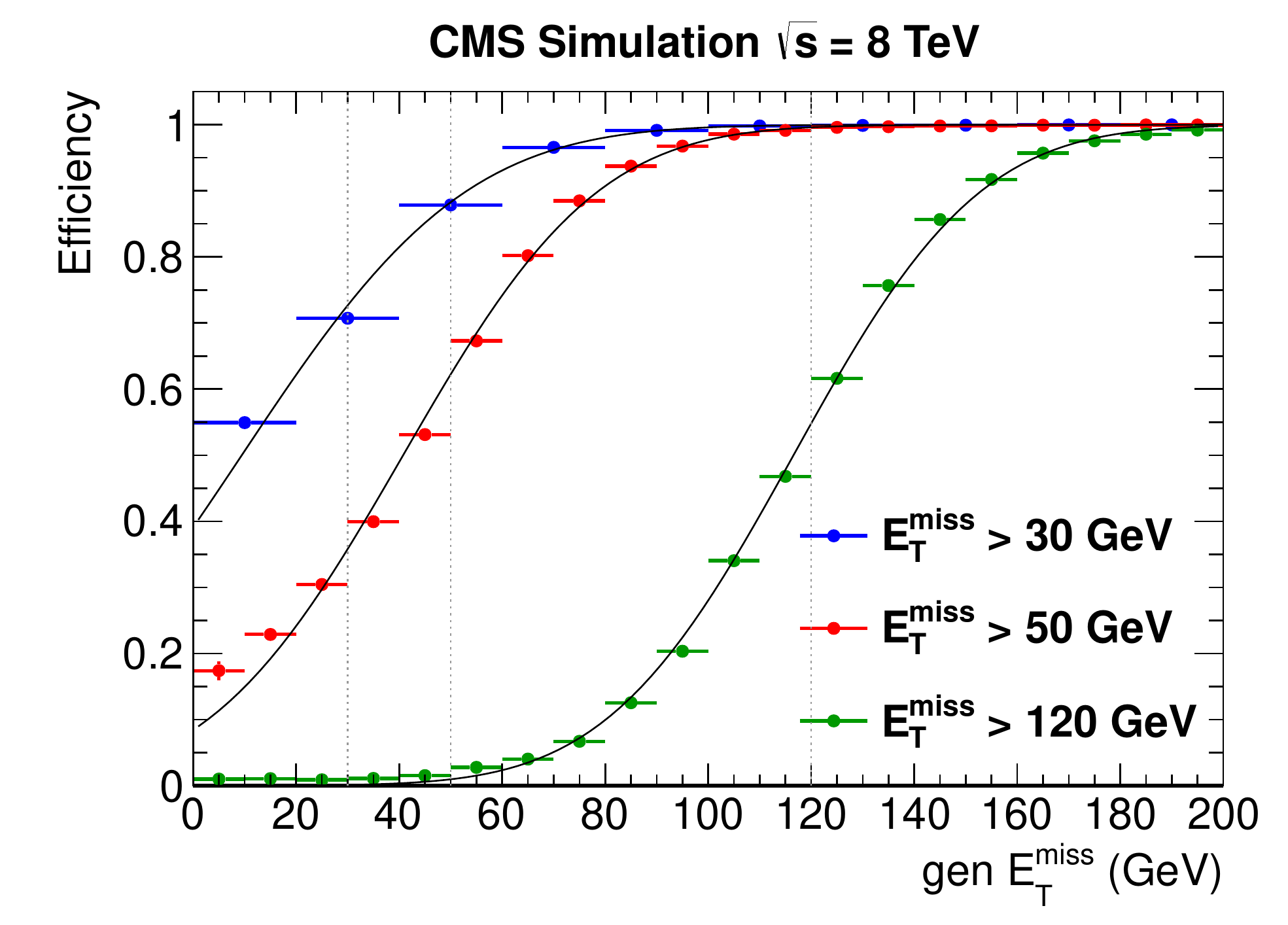}
\includegraphics[width=0.48\linewidth]{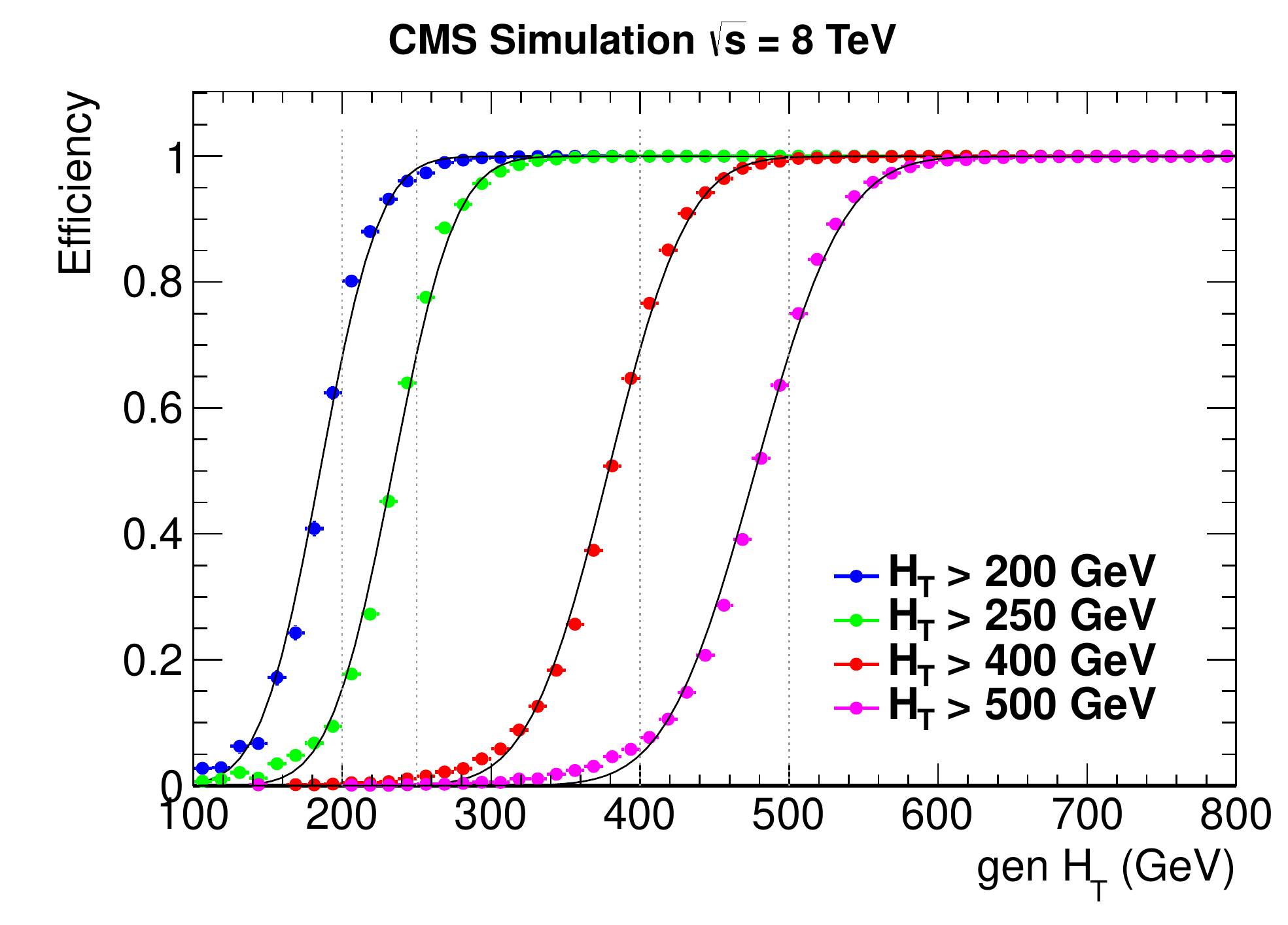}
\end{center}
\caption{\label{fig:htmetThresh}
Efficiency for an event to satisfy a given reconstructed \ETmiss ($\HT$) threshold as a function of generator-level \ETmiss ($\HT^\text{gen}$).
The curves are shown for \ETmiss thresholds of 30, 50, and 120\GeV; the thresholds for \HT\ are 200, 250, 400, and 500\GeV.
}
\end{figure}
The \MET~and \HT turn-on curves, shown in Fig.~\ref{fig:htmetThresh}
as a function of the respective generator-level quantities,
are parametrized as \linebreak[3]$0.5 \cdot \epsilon_{\infty} \cdot \big\{ \erf \big[ ( x - x_{1/2} ) / \sigma\big] + 1 \big\}$,
with $\erf(z)$ the error function, and $\epsilon_{\infty}$, $x_{1/2}$, and $\sigma$ the parameters of the fit.
The generator \HT is calculated using generator jets, obtained by clustering all stable particles from the hard collision, after showering and hadronization, except
for neutrinos and other non-interacting particles.
The parameters of the fitted functions are summarized in Tables~\ref{tab:metThresh} and~\ref{tab:htThresh} for \ETmiss and \HT, respectively.
Analogously to the offline selection, only generator jets that are separated from generator electrons and muons by
$\Delta R \equiv \sqrt{\Delta\phi^2 + \Delta\eta^2} \gt 0.4$ are considered
in the derivation and application of the efficiency model.  Only electrons and muons from the hard collision are considered.
The separation between jets and leptons applies to the calculation of \HT as well as to the counting of jets and b-tagged jets.
The generator-level \ETmiss is constructed as the vector sum $\pt$ of all neutrinos, selected after showering and hadronization, and any other non-interacting
particles from the hard collision.
\begin{table}[h!]
\begin{center}
\topcaption{\label{tab:metThresh} The resulting fit parameters for the efficiency curves presented in Fig.~\ref{fig:htmetThresh} left.}
\begin{tabular}{l|ccc}\hline\hline
Parameter                  &      $\ETmiss>30\GeV$            & $\ETmiss>50$\GeV                 & $\ETmiss>120$\GeV      \\  \hline
$\epsilon_{\infty}$ & $1.000\pm0.001$           & $1.000\pm0.001$           & $0.999\pm0.001$  \\
$x_{1/2}$ (\GeVns{})      & $13.87\pm0.30\phantom{0}$ & $42.97\pm0.14\phantom{0}$ & $117.85\pm0.09\phantom{00}$ \\
$\sigma$ (\GeVns{})       & $42.92\pm0.34\phantom{0}$ & $37.47\pm0.20\phantom{0}$ & $36.90\pm0.14\phantom{0}$ \\
\hline\hline
\end{tabular}
\end{center}
\end{table}
\begin{table}[h!]
\begin{center}
\topcaption{\label{tab:htThresh} The resulting fit parameters for the efficiency curves presented in Fig.~\ref{fig:htmetThresh} right.}
\begin{tabular}{l|cccc}\hline\hline
Parameter                    &       $\HT>200$\GeV          &    $\HT>250$\GeV    &      $\HT>400$\GeV             & $\HT>500$\GeV       \\  \hline
$\epsilon_{\infty}$ & $0.999\pm0.001$           & $0.999\pm0.001$  & $0.999\pm0.001$             & $0.999\pm0.001$  \\
$x_{1/2}$  (\GeVns{})      & $185.2\pm0.4\phantom{00}$ & $233.9\pm0.3\phantom{00}$  & $378.69\pm0.17\phantom{00}$ & $477.3\pm0.2\phantom{00}$  \\
$\sigma$ (\GeVns{})       & $44.5\pm0.6\phantom{0}$   & $46.9\pm0.4\phantom{0}$  & $59.41\pm0.26\phantom{0}$   & $66.05\pm0.25\phantom{0}$  \\
\hline\hline
\end{tabular}
\end{center}
\end{table}
\begin{figure}[t]
\begin{center}
\includegraphics[width=0.48\linewidth]{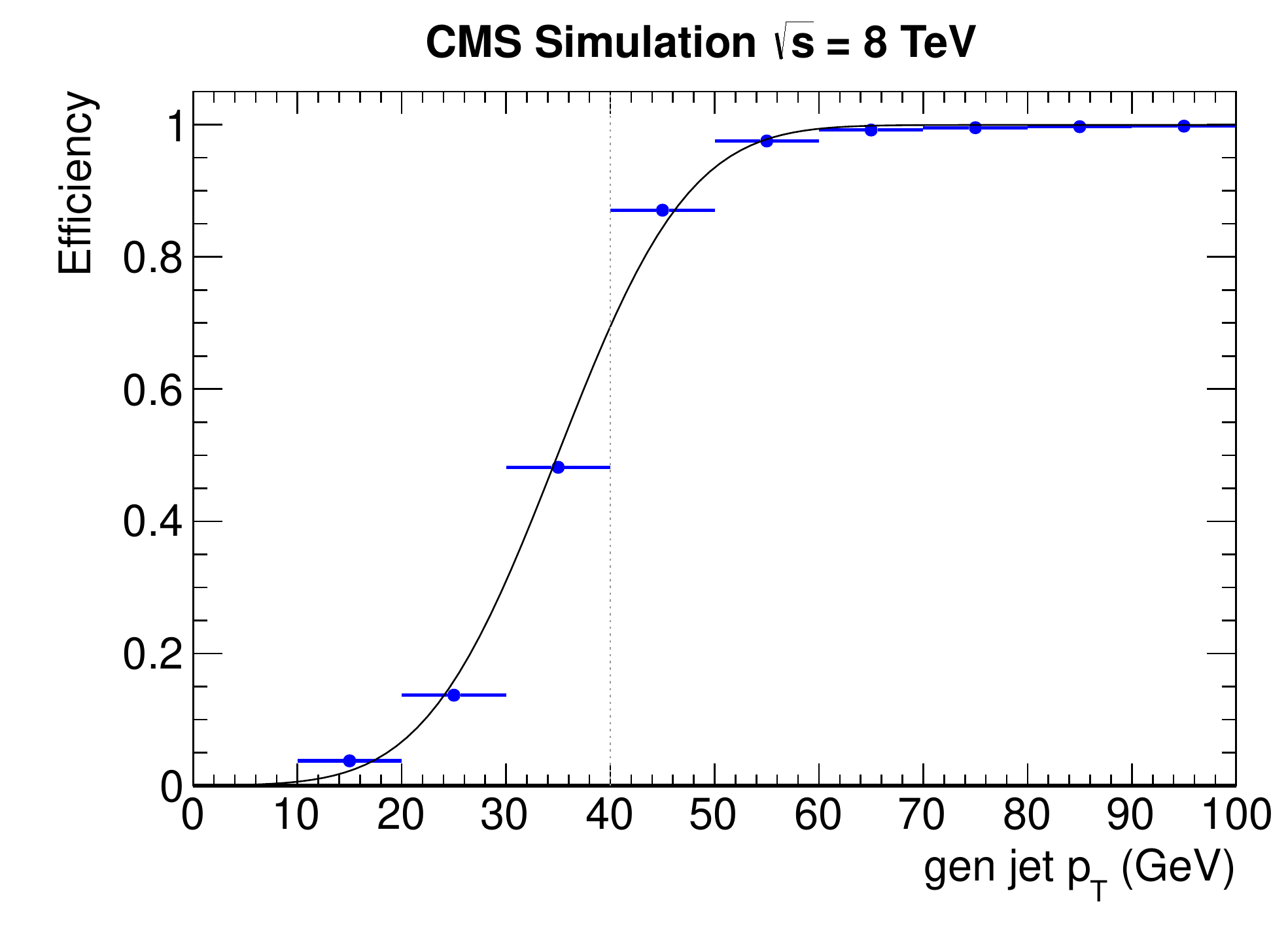}
\includegraphics[width=0.49\linewidth]{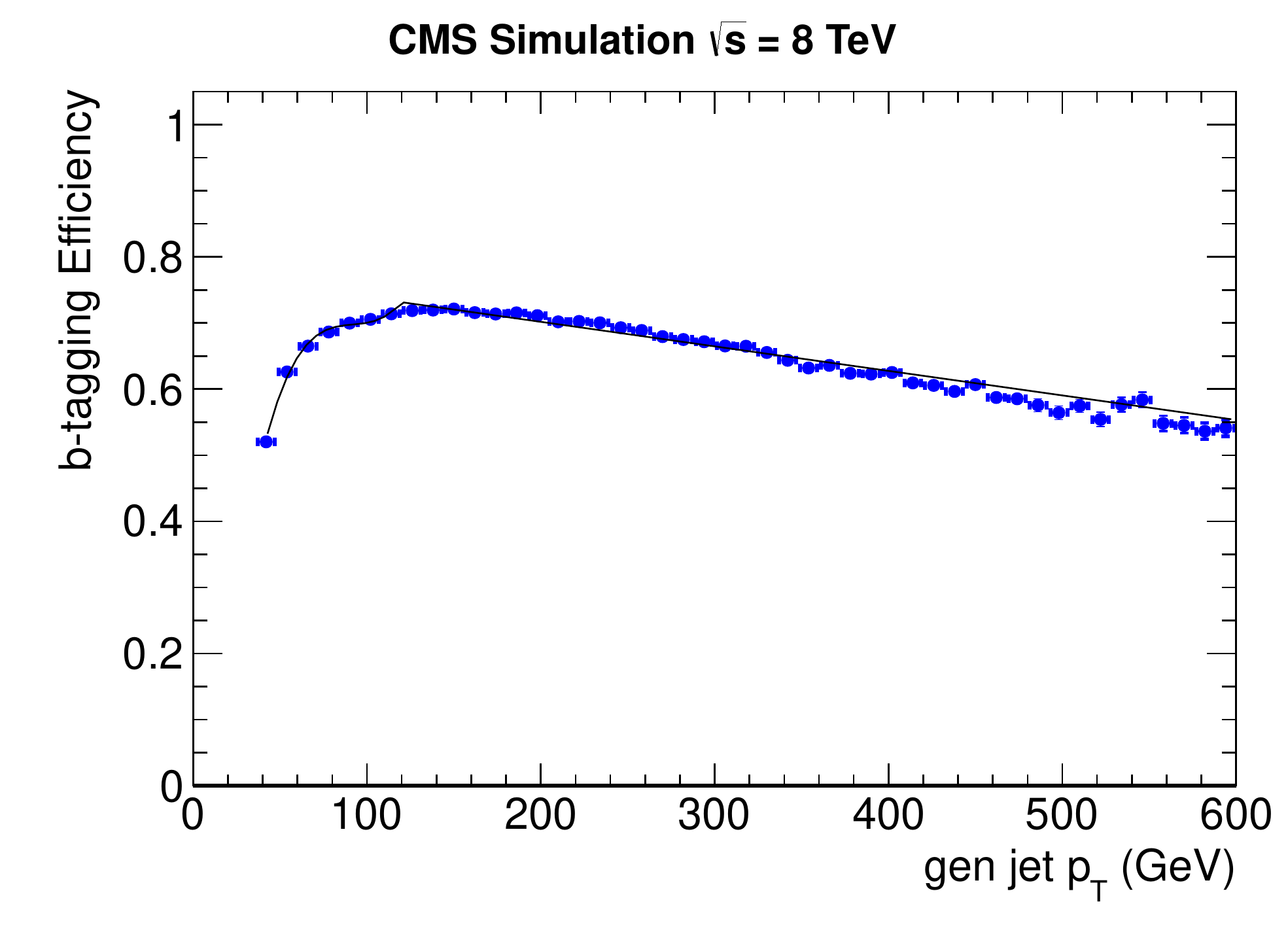}
\caption{\label{fig:jeteffs}
Efficiency for the reconstruction of jets with $\pt > 40$\GeV as a function of the generator jet $\pt$ (left);
b-tagging efficiency as a function of the $\pt$ of the generator jet matched to a bottom quark
from the hard collision (right).
}
\end{center}
\end{figure}

An additional turn-on curve, introduced since the publication of Ref.~\cite{sspaper8TeV10fb}, has been added to parametrize the
efficiency to reconstruct a jet with $\pt > 40$\GeV.  The curve, shown in Fig.~\ref{fig:jeteffs} (left)
as a function of the generator jet $\pt$, is described by the
same functional form as the $\HT$ turn-on.  The parameters of the fit are ($\epsilon_{\infty}$, $x_{1/2}$, $\sigma$)
= (1.0, 29.8\GeV, 18.8\GeV).

Figure~\ref{fig:jeteffs} also shows the b-tagging efficiency,
obtained from simulation, for b quarks with $\abs{\eta} < 2.4$.  The efficiency
is fit with a third-order (first-order) polynomial for $\pt < 120$\GeV ($\pt > 120$\GeV).
The parameters of the fit are given in Table~\ref{tab:befffit}.
\begin{table}[hbt!]
\begin{center}
\topcaption{\label{tab:befffit} b-tagging efficiency parameters.
A polynomial of form $Ax^{3} + Bx^{2} + Cx + D$ is
used for $\pt < 120$\GeV while a linear fit, $Ex + F$, is performed above that threshold.
Note that the parametrization is valid only for moderate range (i.e. [30--600]\GeV) of b-quark jet \pt.}
\begin{tabular}{l|c}
\hline\hline
Parameter & Value \\
\hline
$A$ & $(1.55 \pm 0.05) \times 10^{-6}$  \\
$B$ & $(-4.26 \pm 0.12) \times 10^{-4}$  \\
$C$ & 0.0391 $\pm$ 0.0008 \\
$D$ & $-0.496$  $\pm$ 0.020 \\
$E$ & $(-3.26 \pm 0.01)\times 10^{-4}$ \\
$F$ & 0.7681  $\pm$ 0.0016 \\
\hline\hline
\end{tabular}
\end{center}
\end{table}
\begin{figure}[h]
\begin{center}
\includegraphics[width=0.49\linewidth]{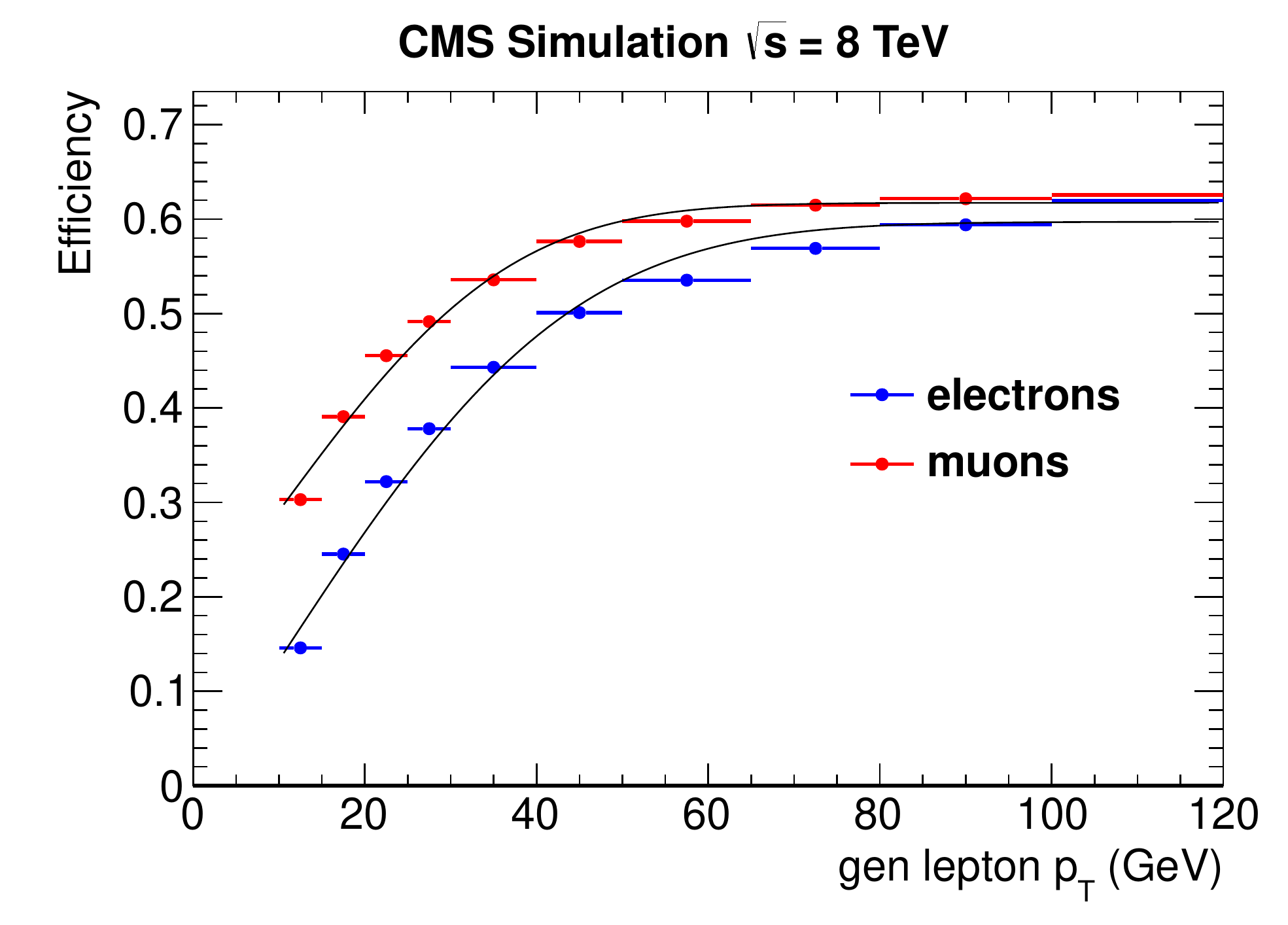}
\end{center}
\caption{\label{fig:lepeff}Electron and muon selection efficiency as a function of the generated lepton \pt.}
\end{figure}

The turn-on curves for the lepton selection are shown in Fig.~\ref{fig:lepeff}.
The lepton efficiency~$(\varepsilon)$---including the effects of reconstruction, identification, and isolation as well as relevant
data-to-simulation scale factors---is parametrized as\linebreak[3]
$\varepsilon(\pt) = \epsilon_{\infty} \cdot \erf \big[ (\pt - 10) / \sigma \big] +
\epsilon_{10} \cdot \big\{ 1 - \erf\big[(\pt - 10) / \sigma\big] \big\}$.
The results of the fit are summarized in Table~\ref{tab:lepefffit}.
\begin{table}[h!]
\begin{center}
\topcaption{\label{tab:lepefffit} The parameters of the fit performed in
Fig.~\ref{fig:lepeff} for electron and muon selection efficiencies.}
\begin{tabular}{l|cc}\hline\hline
Parameter           & Electrons         & Muons             \\ \hline
$\epsilon_{\infty}$ & 0.640 $\pm$ 0.001 & 0.673 $\pm$ 0.001 \\
$\epsilon_{10}$     & 0.170 $\pm$ 0.002 & 0.332 $\pm$ 0.003 \\
$\sigma$ (\GeVns{})        & 36.94 $\pm$ 0.320 & 29.65 $\pm$ 0.382 \\
\hline\hline
\end{tabular}
\end{center}
\end{table}

The prescription to apply the efficiency model is similar to that described in Ref.~\cite{Chatrchyan:2012sa},
with some modifications needed to accommodate the use of exclusive signal regions.  The efficiencies for
the $\HT$ and $\MET$ selections in regions with upper and lower bounds are obtained by taking the
difference between the relevant curves in Fig.~\ref{fig:htmetThresh}.  The jet
reconstruction and b-tagging efficiencies are provided as per-jet quantities.  Thus, one scale factor per
jet should be obtained from the relevant curves.  Additional combinatorial factors should be included,
as dictated by the requirements of the signal region selection.  The application of the lepton efficiency remains
unchanged, with one factor per lepton obtained from the appropriate fit of Fig.~\ref{fig:lepeff}.
All the quoted efficiencies are multiplicative.  The resulting signal yield, obtained
by summing the contribution derived from the efficiency model over all events, is then
compared to the calculated upper limit as described at the beginning of this section.

The efficiency model presented was applied to a variety of the signal models and search regions
considered in this analysis.  Results from the efficiency model were found to agree with those obtained using the
detector simulation and reconstruction to within approximately 30\%.  It should be emphasized that the efficiency
model is approximate and is not universally applicable.  Lepton isolation efficiency, for example, depends on the hadronic
activity in the event and in some extreme cases on the event topology.  For instance, in models giving rise to top
quarks with a significant boost, the lepton isolation efficiency in Fig.~\ref{fig:lepeff} overestimates the true value.

\section{Summary}
\label{sec:conclusion}

We have presented the results of a search for physics beyond the standard model with
same-sign dilepton events using the CMS detector at the LHC.
The study is based on a sample of pp collisions at $\sqrt{s}$ = 8\TeV
corresponding to an integrated luminosity of 19.5\fbinv.
The data are analyzed in exclusive signal regions formed by placing different
requirements on the discriminating variables \HT, \ETmiss, number of jets, and
number of b-tagged jets. The latter can assume values of 0, 1, and 2 or more,
which allow us to probe signatures both with and without third-generation squarks.
No significant deviation from standard model expectation is
observed.

Using sparticle production cross sections calculated in the decoupling limit, and
assuming that gluinos decay exclusively into top or bottom squarks and that
the top and bottom squarks decay as $\sTop_1 \to \cPqt \chiz_1$
and $\sBot_1 \to \cPqt \chim_1$ ($\chim_1 \to \PWm \chiz_1$), lower limits on
gluino and sbottom masses are calculated. Gluinos with masses up to approximately 1050\GeV
and bottom squarks with masses up to about 500\GeV are probed.
In models where gluinos do not decay to third-generation squarks, sensitivity
for gluino masses up to approximately 900\GeV is obtained.
A similar reach in the gluino masses
is demonstrated in the scope of an $R$-parity violating model.

The results are used to set upper limits on the same-sign top-quark pair production cross section
$\sigma(\Pp \Pp \to \cPqt \cPqt,~\cPaqt \cPaqt) < 720$\unit{fb}
and $\sigma(\Pp \Pp \to \cPqt \cPqt) < 370$\unit{fb} at 95\% CL.
An upper limit at 95\% CL of $\sigma(\Pp \Pp \to \cPqt \cPqt \cPaqt \cPaqt) < 49$\unit{fb} is obtained for
the cross section of quadruple top-quark production.

\section*{Acknowledgements}
We wish to congratulate our colleagues in the CERN accelerator departments for the excellent performance of the LHC machine.
We thank the technical and administrative staff at CERN and other CMS institutes, and acknowledge support from:
FMSR (Austria); FNRS and FWO (Belgium); CNPq, CAPES, FAPERJ, and FAPESP (Brazil);
MES (Bulgaria); CERN; CAS, MoST, and NSFC (China); COLCIENCIAS (Colombia);
MSES (Croatia); RPF (Cyprus); Academy of Sciences and NICPB (Estonia); Academy of Finland, MEC, and HIP (Finland);
CEA and CNRS/IN2P3 (France); BMBF, DFG, and HGF (Germany); GSRT (Greece); OTKA and NKTH (Hungary); DAE and DST (India);
IPM (Iran); SFI (Ireland); INFN (Italy); NRF and WCU (Korea); LAS (Lithuania);
CINVESTAV, CONACYT, SEP, and UASLP-FAI (Mexico); MSI (New Zealand); PAEC (Pakistan);
MSHE and NSC (Poland); FCT (Portugal); JINR (Armenia, Belarus, Georgia, Ukraine, Uzbekistan);
MON, RosAtom, RAS and RFBR (Russia); MSTD (Serbia); MICINN and CPAN (Spain);
Swiss Funding Agencies (Switzerland); NSC (Taipei); TUBITAK and TAEK (Turkey); STFC (United Kingdom); DOE and NSF (USA).

Individuals have received support from the Marie-Curie programme and the European Research Council and EPLANET (European Union); the Leventis Foundation; the A. P. Sloan Foundation; the Alexander von Humboldt Foundation; the Belgian Federal Science Policy Office; the Fonds pour la Formation \`a la Recherche dans l'Industrie et dans l'Agriculture (FRIA-Belgium); the Agentschap voor Innovatie door Wetenschap en Technologie (IWT-Belgium); the Ministry of Education, Youth and Sports (MEYS) of Czech Republic; the Council of Science and Industrial Research, India; the Compagnia di San Paolo (Torino); the HOMING PLUS programme of Foundation for Polish Science, cofinanced by EU, Regional Development Fund; and the Thalis and Aristeia programmes cofinanced by EU-ESF and the Greek NSRF.

\bibliography{auto_generated}   

\cleardoublepage \appendix\section{The CMS Collaboration \label{app:collab}}\begin{sloppypar}\hyphenpenalty=5000\widowpenalty=500\clubpenalty=5000\textbf{Yerevan Physics Institute,  Yerevan,  Armenia}\\*[0pt]
S.~Chatrchyan, V.~Khachatryan, A.M.~Sirunyan, A.~Tumasyan
\vskip\cmsinstskip
\textbf{Institut f\"{u}r Hochenergiephysik der OeAW,  Wien,  Austria}\\*[0pt]
W.~Adam, T.~Bergauer, M.~Dragicevic, J.~Er\"{o}, C.~Fabjan\cmsAuthorMark{1}, M.~Friedl, R.~Fr\"{u}hwirth\cmsAuthorMark{1}, V.M.~Ghete, C.~Hartl, N.~H\"{o}rmann, J.~Hrubec, M.~Jeitler\cmsAuthorMark{1}, W.~Kiesenhofer, V.~Kn\"{u}nz, M.~Krammer\cmsAuthorMark{1}, I.~Kr\"{a}tschmer, D.~Liko, I.~Mikulec, D.~Rabady\cmsAuthorMark{2}, B.~Rahbaran, H.~Rohringer, R.~Sch\"{o}fbeck, J.~Strauss, A.~Taurok, W.~Treberer-Treberspurg, W.~Waltenberger, C.-E.~Wulz\cmsAuthorMark{1}
% ...
\vskip\cmsinstskip
% ...
1:~~Also at Vienna University of Technology, Vienna, Austria\\
2:~~Also at CERN, European Organization for Nuclear Research, Geneva, Switzerland\\
% ...

\end{sloppypar}
\end{document}